\documentclass[12pt]{iopart}
\RequirePackage{lineno} 

\usepackage{verbatim}
\usepackage{cite} 
\usepackage{cancel}
\usepackage{harpoon}
\usepackage{graphicx}
\usepackage[colorlinks,linkcolor=blue,citecolor=blue,filecolor=black,urlcolor=blue]{hyperref}

\def\st{\begin{eqnarray}}
\def\stp{\end{eqnarray}}
\usepackage{color}
\usepackage{harpoon}
\definecolor{my}{rgb}{1, 0, 0}

\newcommand{\Deta}{\mbox{$\Delta \eta$}}

\newcommand{\pT}{\mbox{$p_{\mathrm{T}}$}}
\newcommand{\eT}{E_{\mathrm{T}}}
\newcommand{\sqrtnn}{\sqrt{s_{\mathrm{nn}}}}

\newcommand{\npart}{\mbox{$N_{\mathrm{part}}$}}
\newcommand{\npartf}{N_{\mathrm {part}}^{\mathrm{F}}}
\newcommand{\npartb}{N_{\mathrm {part}}^{\mathrm{B}}}
\begin{document}

\title[Event-by-Event flow]{Event-shape fluctuations and flow correlations in ultra-relativistic heavy-ion collisions}

\author{Jiangyong Jia}

\address{Department of Chemistry, Stony Brook University, Stony Brook, NY 11794, USA,}
\address{Physics Department, Brookhaven National Laboratory, Upton, NY 11796, USA}
\ead{jjia@bnl.gov}
\begin{abstract}
I review recent measurements of a large set of flow observables associated with event-shape fluctuations and collective expansion in heavy ion collisions. First, these flow observables are classified and experiment methods are introduced. The experimental results for each type of observables are then presented and compared to theoretical calculations. A coherent picture of initial condition and collective flow based on linear and non-linear hydrodynamic responses is derived, which qualitatively describe most experimental results. I discuss new types of fluctuation measurements that can further our understanding of the event-shape fluctuations and collective expansion dynamics.
\end{abstract}
\submitto{\JPG}


\maketitle
\section{Introduction}
Relativistic heavy ion collisions at the RHIC and the LHC create a hot and dense nuclear matter that is composed of strongly interacting quarks and gluons. This initially produced matter has an asymmetric shape in the transverse plane. Driven by the large pressure gradients arising from the strong interactions, the matter expands collectively and transfers the asymmetry in the initial geometry into azimuthal anisotropy of produced particles in momentum space~\cite{Ollitrault:1992bk,Alver:2010gr}. Hydrodynamic models are used to understand the space-time evolution of the matter from the measured azimuthal anisotropy. The success of these models in describing the anisotropy of particle production in heavy-ion collisions at RHIC and the LHC~\cite{Adare:2011tg,Adamczyk:2013waa,ALICE:2011ab,Aad:2012bu,Chatrchyan:2013kba,Aad:2013xma,Aad:2014fla} places important constraints on the transport properties, such as ratio of shear viscosity to entropy density $\eta/s$, and initial conditions of the produced matter~\cite{Luzum:2012wu,Teaney:2010vd,Gale:2012rq,Niemi:2012aj,Qiu:2012uy,Teaney:2013dta}.

For many years, the initially-produced fireball was treated as a smooth and boost invariant distribution of quarks and gluons, given by the overlap of two Woods-saxon functions that describe the distribution of nucleons in the two colliding nuclei. One important insight emerged around 2010 is the dominant role of event-by-event (EbyE) fluctuation~\cite{Alver:2010gr} of nuclear wave-functions: the transverse positions of nucleons in the overlap region, as well as the parton density profiles inside those nucleons can fluctuate from collision to collision (earlier work on fluctuations can be found in Ref.~\cite{Miller:2003kd}). As a result, each collision produces a different and lumpy fireball, each has its own shape and each features its own hydrodynamic expansion. Indeed, the azimuthal distribution of produced particles, when expanded into a Fourier series, shows significant harmonics up to at least 6$^{\mathrm th}$ order. The relative strength of these harmonics, their centrality, transverse momentum ($\pT$) and particle mass dependence are well described by hydrodynamic calculations that were also performed on an EbyE basis~\cite{Alver:2010dn,Gale:2012in,Gardim:2012yp,Song:2013qma}.

The rich patterns in the initial state fluctuations and the resulting hydrodynamic expansion also imply a large space of information associated with many new flow observables~\cite{Bhalerao:2011yg,Qin:2011uw,Jia:2012ma,Huo:2013qma,Jia:2014ysa}. Initial measurements have been carried out on some of these observables, noticeably the probability distribution of individual harmonics~\cite{Aad:2013xma} and amplitude or phase correlations between different harmonics~\cite{ATLAS2014-022,Aad:2014fla}. Studies of these observables are further augmented using recently proposed event-shape techniques~\cite{Schukraft:2012ah,Huo:2013qma,Jia:2014vja}. These measurements already provided unprecedented insights on the nature of the initial density fluctuations and dynamics of the collective evolution. Future detailed mapping of these flow observables requires concerted efforts. The goal of this article is to provide a concise introduction of these new flow observables, review the status of current measurements, and discuss open issues and future physics opportunities in exploring these observables. More information on the current status of the harmonic flow measurements can be found in a contribution by R.~Snellings in the same issue~\cite{snellings}.

\section{Event-by-event flow observables}
\label{sec:2}
\subsection{Eccentricities and azimuthal flow harmonics}
\label{sec:21}
When describing the transverse expansion dynamics, it is convenient to parameterize the probability distribution of the particle production in azimuthal angle $\phi$ in each event by a Fourier expansion:
\begin{equation}
\label{eq:211}
dN/d\phi\propto1+2\sum_{n=1}^{\infty}v_{n}\cos n(\phi-\Phi_{n})\;,
\end{equation}
where $v_n$ and $\Phi_n$ represent the magnitude and the phase (referred to as the event plane or EP) of the $n^{\mathrm{th}}$-order harmonic flow, which are often represented as a two-dimensional vector or in a complex form: 
\begin{eqnarray}
\label{eq:212} 
\vec{v}_n=(v_n\cos n\Phi_{n}, v_n\sin n\Phi_{n}) \equiv v_n e^{i n\Phi_n}.
\end{eqnarray} 
The first few harmonics are referred to as dipolar, elliptic, triangular, quadrangular flow etc. 

Since the number of particles in each event is finite, the values of $v_n$ and $\Phi_n$ can not be obtained event-by-event. Instead, they are estimated using the azimuthal distribution of particles in the event:
\begin{eqnarray}
\label{eq:213} 
\vec{v}_n^{\;\mathrm{obs}}=(v_n^{\mathrm{obs}}\cos n\Psi_{n}, v_n^{\mathrm{obs}}\sin n\Psi_{n})\equiv v_n^{\mathrm{obs}} e^{i n\Psi_n}= \left\langle e^{i n\phi} \right\rangle\;.
\end{eqnarray} 
where the average is over all produced particles in the event. The estimated flow vector $\vec{v}_n^{\;\mathrm{obs}}$ smears around $\vec{v}_n$. This smearing can be removed statistically via an unfolding method~\cite{Aad:2013xma} or corrected for {\it on average} via the event-plane method~\cite{Ollitrault:1992bk} or the scalar-product method~\cite{Adler:2002pu}.

The development of harmonic flow is driven by the asymmetries in the pressure gradients of the matter, which in turn is controlled by the detailed shape configuration of the initial density profile. The shape configuration of each event is often characterized by a set of eccentricity vector $\vec{\varepsilon}_n$, calculated from the transverse positions $(r,\phi)$ of the participating nucleons relative to their center of mass~\cite{Alver:2010gr,Teaney:2010vd}:
\begin{eqnarray}
\label{eq:214}
\hspace*{-1.5cm}\vec{\varepsilon}_n= (\varepsilon_n\cos n\Phi_{n}^*,\varepsilon_n\sin n\Phi_{n}^*) \;\;\mathrm{or} \;\; \varepsilon_n e^{i n\Phi_n^*} = -\frac{\langle r^m e^{i n\phi} \rangle}{\langle r^m\rangle}\;, m=\left\{\begin{array}{ll} 3 & n=1\\
n & n>1\\   \end{array}\right.
\end{eqnarray} 
where $\langle...\rangle$ denotes an average over the transverse position of all participating nucleons, and the $\varepsilon_n$ and angle $\Phi_n^*$ (also known as participant-plane, PP) represent the magnitude and orientation of the eccentricity vector, respectively. Hydrodynamic calculations shows that the first few flow harmonics are directly related to the eccentricities of the corresponding order, e.g. $\vec{v}_n \propto \vec{\varepsilon}_n$ for $n\leq3$~\cite{Qiu:2011iv,Gardim:2011xv}. By default the radial weights are chosen as $m=n$ for $n>1$, and $m=3$ for $n=1$~\cite{Teaney:2010vd}. But sometimes they are also calculated using other weights, for example $m=2$~\cite{Alver:2010gr}. These alternative definitions of eccentricity capture the degree of freedom along the radial direction, and reflect important information of the system, such as the system size and density gradients~\cite{Teaney:2010vd,Gardim:2011xv,Floerchinger:2014fta}. Events with the same $\varepsilon_n$ values in the default definition Eq.~\ref{eq:214} may have different values when calculated with alternative weights.


\subsection{Flow fluctuations and observables}
\label{sec:22}

In heavy ion collisions, the number of participating nucleons $\npart$ is finite and their positions fluctuate randomly in the transverse plane, leading to strong EbyE fluctuations of $\varepsilon_n$ and $\Phi_n^*$. These fluctuations also result in non-trivial correlations between eccentricities and PP angles of different order characterized by $p(\varepsilon_n,\varepsilon_m,...,\Phi_n^*,\Phi_m^*,...)$~\cite{Huo:2013qma,Jia:2014ysa}. Consequently, the matter created in each collision follows a different collective expansion, with its own set of flow harmonics. Experimental observables describing harmonic flow can be generally given by the joint probability distribution (pdf) of all $v_n$ and $\Phi_n$:
\begin{equation}
\label{eq:220}
p(v_n,v_m,...., \Phi_n, \Phi_m, ....)=\frac{1}{N_{\mathrm{evts}}}\frac{dN_{\mathrm{evts}}}{dv_ndv_m...d\Phi_{n}d\Phi_{m}...},
\end{equation}
with each variable being a function of $\pT$, $\eta$ etc~\cite{Gardim:2012im}. Among these, the joint probability distribution of the EP angles
\begin{eqnarray}
\nonumber
\frac{dN_{\mathrm{evts}}}{d\Phi_{1}d\Phi_{2}...d\Phi_{l}} &\propto& \sum_{c_n=-\infty}^{\infty} a_{c_1,c_2,...,c_l} \cos(c_1\Phi_1+c_2\Phi_2...+c_l\Phi_l),\\\label{eq:221}
a_{c_1,c_2,...,c_l}&=&\left\langle\cos(c_1\Phi_1+c_2\Phi_2+...+c_l\Phi_l)\right\rangle
\end{eqnarray}
can be reduced to the following EP correlators~\cite{Bhalerao:2011yg,Qin:2011uw,Jia:2012ma}:
\begin{eqnarray}
\label{eq:222}
\left\langle\cos(c_1\Phi_1+2c_2\Phi_2...+lc_l\Phi_l)\right\rangle, c_1+2c_2...+lc_l=0.
\end{eqnarray}

Due to $n$-fold symmetry of $\Phi_n$, these correlators should be invariant under a phase shift $\Phi_n\rightarrow \Phi_n+2\pi/n$. It should also be invariant under a global rotation by any angle. The first condition requires that the EP angle of the $n^{\mathrm{th}}$-order harmonic appears as integer multiple of $n\Phi_n$, while the second condition requires the sum of the coefficients to vanish. Together they lead to the constraint in Eq.~\ref{eq:222}.

Heavy ion experiments at RHIC and LHC have recorded billions of Pb+Pb or Au+Au collisions, which are distributed according to the underlying pdf given by Eq.~\ref{eq:220}. However, it is more practical to measure the projections of the full probability distribution on a finite number of variables. These projected distributions can be generally classified into three types: 1) those involving only the flow amplitudes, 2) those involving only the flow phases or event-plane angles, and 3) those involving both amplitudes and phases. These are listed in the left column of Tab.~\ref{tab:1}.


It is not always straightforward to directly measure the flow probability distributions, instead, the information of Eq.~\ref{eq:220} can also be obtained in terms of its moments measured by a $m$-particle azimuthal correlation of the following general form~\cite{Bhalerao:2011yg}:
\small{\begin{eqnarray}\nonumber
\hspace*{-2cm}\left\langle\left\langle e^{in_1\phi_1}e^{in_2\phi_2}...e^{in_m\phi_m}\right\rangle\right\rangle &=&\left\langle v_{n_1}^{\mathrm{obs}}e^{in_1\Psi_1}\;v_{n_2}^{\mathrm{obs}}e^{in_2\Phi_2}\;...\;v_{n_m}^{\mathrm{obs}}e^{in_m\Psi_m}\right\rangle\\\nonumber
&=&\left\langle v_{n_1}e^{in_1\Phi_1}\;v_{n_2}e^{in_2\Phi_2}\;...\;v_{n_m}e^{in_m\Phi_m}\right\rangle+\mbox{\small{non-flow}\normalsize}\\\label{eq:223b}
&=&\left\langle v_{n_1}v_{n_2}...v_{n_m} \cos(n_1\Phi_{n_1}+n_2\Phi_{n_2}...+n_m\Phi_{n_m})\right\rangle+\mbox{\small{non-flow}\normalsize}
\end{eqnarray}}\normalsize
where $\Sigma n_i=0$. The double average in the left hand side is carried out over all $m$-particles in one event then over all events, while the single averages are carried out over the events. All sine terms drops out after the averages. The event average also removes any statistical smearing effects, leading to the second line of the equation. Note that the angle $\phi_n$ refers to the azimuthal angle of $n^{\mathrm{th}}$ particle, while $\Psi_{n}$ and $\Phi_{n}$ refers to the $n^{\mathrm{th}}$-order observed and true event plane, respectively. This equation is also conveniently expressed as:
\small{\begin{eqnarray}
\nonumber
\hspace*{-2cm}\left\langle \cos(n_1\phi_1+n_2\phi_2...+n_m\phi_m)\right\rangle=\left\langle v_{n_1}v_{n_2}...v_{n_m} \cos(n_1\Phi_{n_1}+n_2\Phi_{n_2}...+n_m\Phi_{n_m})\right\rangle+\mbox{non-flow}\\\label{eq:223}
\end{eqnarray}}\normalsize
In the absence of non-flow, the $m$-particle correlation reduces to one particular $m^{\mathrm{th}}$-order moment (Eq.~\ref{eq:223}) of the underlying flow probability distribution.

According to the traditional definition of moments or cumulants, the $m^{\mathrm{th}}$-order moment of the underlying pdf Eq.~\ref{eq:220} should has the form of $\langle X_{n_1}X_{n_2}...X_{n_m}\rangle$ with $X_{k} = v_{k}$ or $\Phi_{k}$. However the azimuthal correlation analysis discussed above uses a different definition $X_{k} = v_{k}e^{ik\Phi_{k}}$, which in general does not capture the full information of the pdf. For example, since $\left\langle X_{k}\right\rangle=0$,  all the odd moments of $p(v_n)$, such as $\left\langle v_n\right\rangle$ or $\left\langle v_n^3\right\rangle$, can not be accessed directly via the multi-particle correlation method. 

The $m$-particle correlation is often combined with correlations involving less number of particles to construct the corresponding $m$-particle cumulant, which removes non-flow correlations of order less than $m$, i.e:
\begin{eqnarray}
\left\langle X_{n_1}X_{n_2}...X_{n_m}\right\rangle_c = \left\langle X_{n_1}X_{n_2}...X_{n_m}\right\rangle - \mbox{all\;lower-order correlations}
\end{eqnarray}
For example, $\left\langle XY\right\rangle_c = \left\langle XY\right\rangle - \left\langle X\right\rangle \left\langle Y\right\rangle$, and $\left\langle XYZ\right\rangle_c = \left\langle XYZ\right\rangle - \left\langle XY\right\rangle \left\langle Z\right\rangle-\left\langle YZ\right\rangle \left\langle Z\right\rangle-\left\langle ZX\right\rangle \left\langle Y\right\rangle+2\left\langle X\right\rangle \left\langle Y\right\rangle\left\langle Z\right\rangle$. However, since $X_{k} = v_{k}e^{ik\Phi_{k}}$, many combinations vanish in multi-particle correlation analysis (see some concrete examples are given below).

Each of the three types of reduced pdfs in Tab.~\ref{tab:1} has its own multi-particle correlations from Eq.~\ref{eq:223}. The corresponding cumulant involves particular combination with several lower order moments. For example, ignoring non-flow, the first few moments and cumulants of $p(v_n)$ accessible to the correlation analysis are :
\st\nonumber
\hspace*{-1.5cm}\left\langle\left\langle \cos (n\phi_1-n\phi_2)\right\rangle\right\rangle &=& \left\langle v_n^2\cos (n\Phi_n-n\Phi_n) \right\rangle = \left\langle v_n^2 \right\rangle\\\nonumber
\hspace*{-1.5cm}\left\langle\left\langle \cos (n\phi_1-n\phi_2+\textcolor{red}{n\phi_3-n\phi_4})\right\rangle\right\rangle &=& \left\langle v_n^4\cos (n\Phi_n-n\Phi_n+\textcolor{red}{n\Phi_n-n\Phi_n}) \right\rangle=\left\langle v_n^4 \right\rangle\\
\hspace*{-1.cm}...
\stp
\st\nonumber
\hspace*{-1.5cm}\left\langle\left\langle \cos (n\phi_1-n\phi_2)\right\rangle\right\rangle_c &=& \left\langle v_n^2 \right\rangle\\\nonumber
\hspace*{-1.5cm}\left\langle\left\langle \cos (n\phi_1-n\phi_2+\textcolor{red}{n\phi_3-n\phi_4})\right\rangle\right\rangle_c&=&\left\langle \cos (n\phi_1-n\phi_2+\textcolor{red}{n\phi_3-n\phi_4})\right\rangle-\\\nonumber
&& \hspace*{-5cm}         \left\langle\left\langle \cos (n\phi_1-n\phi_2) \right\rangle\right\rangle\left\langle\left\langle \cos (\textcolor{red}{n\phi_3-n\phi_4}) \right\rangle\right\rangle-\left\langle\left\langle \cos (n\phi_1-\textcolor{red}{n\phi_4}) \right\rangle\right\rangle\left\langle\left\langle \cos (n\phi_2-\textcolor{red}{n\phi_3}) \right\rangle\right\rangle\\\nonumber&=&\left\langle v_n^4 \right\rangle-2\left\langle v_n^2 \right\rangle^2\\\label{eq:pmix1}
\hspace*{-1.cm}...
\stp
with subscript ``$c$'' denoting the cumulant form and with the assumption that $\Phi_n$ is the same for the two particles. The single particle flow coefficients $v_n\{2k\}$ are then calculated from these cumulants, e.g:
\st\nonumber
v_n\{2\}^2\equiv\langle v_n^2\rangle\;, -v_n\{4\}^4\equiv\langle v_n^4\rangle-2\langle v_n^2\rangle^2\;,\;...
\stp

Similarly, the lowest-order moment and cumulant for $p(v_n,v_m)$, which are accessible to the correlation analysis, involve four-particle correlation of the following form:
\st\nonumber
\hspace*{-2.3cm}\left\langle\left\langle \cos (n\phi_1-n\phi_2+\textcolor{red}{m\phi_3-m\phi_4})\right\rangle\right\rangle &=& \left\langle v_n^2v_m^2\cos (n\Phi_n-n\Phi_n+\textcolor{red}{m\Phi_m-m\Phi_m}) \right\rangle = \left\langle v_n^2 \textcolor{red}{v_m^2} \right\rangle\\\nonumber
\hspace*{-2.3cm}\left\langle\left\langle \cos (n\phi_1-n\phi_2+\textcolor{red}{m\phi_3-m\phi_4})\right\rangle\right\rangle_c &=& 
\left\langle\left\langle \cos (n\phi_1-n\phi_2+\textcolor{red}{m\phi_3-m\phi_4})\right\rangle\right\rangle - \\\nonumber 
&&\hspace*{-2.3cm} \left\langle\left\langle \cos (n\phi_1-n\phi_2)\right\rangle\right\rangle\left\langle\left\langle \cos (\textcolor{red}{m\phi_3-m\phi_4})\right\rangle\right\rangle = \left\langle v_n^2 \textcolor{red}{v_m^2} \right\rangle-\left\langle v_n^2 \right\rangle\left\langle \textcolor{red}{v_m^2} \right\rangle\\
\stp
Clearly, this four-particle cumulant, first proposed in Ref.~\cite{Bilandzic:2013kga}, reduces to zero if flow magnitudes $v_n$ and $v_m$ are un-correlated. Similar relations can be easily derived for correlation between three or more flow magnitudes. 

\newcommand{\sgn}{\mathrm{sgn}}
The correlation between event-plane angles, $p(\Phi_n,\Phi_m,...)$, can be accessed via multi-particle correlation derived from Eq.~\ref{eq:223} ($\sgn(x)$ denotes the sign of $x$):
\small{\begin{eqnarray}
\nonumber
\hspace*{-2cm}\left\langle\left\langle \cos (\sgn(c_1)\Sigma_{i_1=1}^{|c_1|}\phi_{i_1}+...+\sgn(c_l)\Sigma_{i_l=1}^{|c_l|}l\phi_{i_l})\right\rangle\right\rangle=\left\langle v_1^{|c_1|}...v_l^{|c_l|}\cos(c_1\Phi_1+...+lc_l\Phi_l)\right\rangle\\
\end{eqnarray}}\normalsize
which involves $\Sigma_{i=1}^{l} |c_i|$ number of particles with $\Sigma_{i=1}^{l} (ic_i) = 0$. The corresponding cumulant has identical expression. The right-hand side of this expression, referred to as the scalar product (SP)~\cite{Adler:2002pu,Luzum:2012da}, is similar to that defined in Eq.~\ref{eq:222}, except for the $v_n$ weight. This definition was argued to be preferable over Eq.~\ref{eq:222} as the results are independent of resolution of the event planes in the experiments~\cite{Luzum:2012da,Bhalerao:2013ina}.

The last category of pdf in Tab.~\ref{tab:1} is the mixed correlations involving both magnitudes and the phases of harmonic flow, e.~g. $p(v_l,\Phi_n,\Phi_m,...)$. The cumulant form can be easily obtained from Eq.~\ref{eq:223}. For example, the lowest-order moment and cumulant for $p(v_2,\Phi_3,\Phi_6)$ accessible to the correlation analysis can be obtained from a five-particle correlation:
\small{\st\nonumber
\hspace*{-2.4cm} \left\langle\left\langle \cos (3\phi_1+3\phi_2-6\phi_3+\textcolor{red}{2\phi_4-2\phi_5})\right\rangle\right\rangle&=&\left\langle \textcolor{red}{v_2^2}v_3^2v_6\cos(3\Phi_3+3\Phi_3-6\Phi_6+\textcolor{red}{2\Phi_2-2\Phi_2})\right\rangle\\
\hspace*{-2.4cm}&=&\left\langle \textcolor{red}{v_2^2}v_3^2v_6\cos6(\Phi_3-\Phi_6)\right\rangle\\\nonumber
\hspace*{-2.4cm}\left\langle\left\langle \cos (3\phi_1+3\phi_2-6\phi_3+\textcolor{red}{2\phi_4-2\phi_5})\right\rangle\right\rangle_c&=&\left\langle \textcolor{red}{v_2^2}v_3^2v_6\cos6(\Phi_3-\Phi_6)\right\rangle-\left\langle \textcolor{red}{v_2^2}\right\rangle\left\langle v_3^2v_6\cos6(\Phi_3-\Phi_6)\right\rangle\\
\stp}\normalsize
This cumulant reduces to zero if flow amplitudes and EP angles are un-correlated. Note that if some of the indices are the same, the cumulants takes somewhat different form. For example the lowest-order non-zero cumulants for $p(v_2,\Phi_2,\Phi_4)$ is:
\small{\st\nonumber
\hspace*{-2.4cm}\left\langle\left\langle \cos (2\phi_1+2\phi_2-4\phi_3+2\phi_4-2\phi_5)\right\rangle\right\rangle_c&=&\left\langle v_2^4v_4\cos4(\Phi_2-\Phi_4)\right\rangle-3\left\langle v_2^2\right\rangle\left\langle v_2^2v_4\cos4(\Phi_2-\Phi_4)\right\rangle\\\label{eq:pmix2}
\stp}\normalsize
A factor of three in the second term in the right-hand of the equation arises because one can swap $\phi_1$ or $\phi_2$ with $\phi_4$.

The flow observables listed in Tab.~\ref{tab:1} can also be accessed via the recently proposed event-shape selection method~\cite{Schukraft:2012ah,Huo:2013qma,Jia:2014vja,ATLAS2014-022}. In this method, events in a narrow centrality interval are further classified according to the observed $v_m$ signal ($m=2$ and 3) in a forward rapidity range. This classification selects events with similar multiplicity but very different ellipticity or triangularity. The values of $v_n$ are then measured at mid-rapidity using the standard flow techniques. Since the $p(v_m)$ distributions are very broad, the correlation of $v_m$ with $v_n$ and/or $\Phi_l$ can be explored over a wide $v_m$ range. The event-shape selection techniques are also sensitive to any differential correlation between $\varepsilon_m$ and $\varepsilon_n$ for fixed centrality, which would otherwise be washed-out when averaging over different initial configurations. One example is the strong anti-correlation between $\varepsilon_2$ and $\varepsilon_3$ predicted by the MC Glauber model~\cite{Lacey:2013eia,Huo:2013qma}. A recent transport model calculation shows that this correlation survives the collective expansion and appears as a similar anti-correlation between $v_2$ and $v_3$~\cite{Huo:2013qma}.

\newcommand{\specialcell}[2][l]{%
\begin{tabular}[#1]{@{}l@{}}#2\end{tabular}}
\begin{table}[!h]
\centering\small{
\begin{tabular}{c|c|c|c}\hline
                         & pdfs &cumulants & event-shape method \tabularnewline\hline
                         & $p(v_n)$      &$v_n\{2k\}$, $k=1$,2,...& \tabularnewline\cline{2-3}
                         &      && \tabularnewline
                         & $p(v_n,v_m)$  &$\langle v_n^2v_m^2\rangle-\langle v_n^2\rangle\langle v_m^2\rangle$, $n\neq m$ &\tabularnewline
                         &      &...& \tabularnewline\cline{2-3}
\specialcell[2]{Flow-\\amplitudes}    & $p(v_n,v_m,v_l)$  &\specialcell[4]{\\\;\;\;\;\;$\langle v_n^2v_m^2v_l^2\rangle+2\langle v_n^2\rangle\langle v_m^2\rangle\langle v_l^2\rangle-$\\$\langle v_n^2v_m^2\rangle\langle v_l^2\rangle-\langle v_m^2v_l^2\rangle\langle v_n^2\rangle-\langle v_l^2v_n^2\rangle\langle v_m^2\rangle$} &yes \tabularnewline
                         &      & $n\neq m \neq l$& \tabularnewline
                         &      & ... & \tabularnewline\cline{2-3}
                         & ...  &Obtained recursively as above& \tabularnewline\hline
\specialcell[2]{EP-\\correlation}   & $p(\Phi_n,\Phi_m,...)$&\specialcell[2]{\\$\langle v_n^{|c_n|}v_m^{|c_m|}...\cos (c_nn\Phi_{n}+c_mm\Phi_{m}+...)\rangle$\\\;\;\;\;\;\;\;\;\;\;\;\;\;\;\;\;\;\;\;\;$\sum_kkc_k=0$}&yes \tabularnewline\hline
\specialcell[2]{Mixed-\\correlation}       & $p(v_l,\Phi_n,\Phi_m,...)$&\specialcell[3]{\\$\langle v_l^2 v_n^{|c_n|}v_m^{|c_m|}...\cos (c_nn\Phi_{n}+c_mm\Phi_{m}+...)\rangle-$\\$\langle v_l^2 \rangle \langle v_n^{|c_n|}v_m^{|c_m|}...\cos (c_nn\Phi_{n}+c_mm\Phi_{m}+...)\rangle$\\\;\;\;\;\;\;\;\;\;\;\;\;\;\;\;$\sum_kkc_k=0$, $n\neq m \neq l$...}&yes \tabularnewline\hline
\end{tabular}}\normalsize
\caption{\label{tab:1} Event-by-event flow observables in terms of probability density distributions (left column) and lowest-order cumulants accessible to the correlation analyses (middle column). Most of them can also be accessed via event-shape selection methods (right column). Note that if some of the indices are the same, the cumulants take somewhat different form, e.g. Eq.~\ref{eq:pmix2}.}
\end{table}

Last few years witnessed impressive progresses in studying these flow observables, both experimentally and theoretically. They have greatly improved our understanding of the fluctuations in the initial density profile and hydrodynamics response in the final state~\cite{Gale:2013da,Heinz:2013th,Luzum:2013yya}. We now know that the elliptic flow and triangular flow are the dominant harmonics, and they are driven mainly by the linear response to the ellipticity and triangularity of the initially produced fireball~\cite{Qiu:2011iv,Gardim:2011xv}:
\begin{eqnarray}
\label{eq:224}
v_2e^{i2\Phi_2} \propto \varepsilon_2e^{i2\Phi^*_2},\;\; v_3e^{i3\Phi_3} \propto \varepsilon_3e^{i3\Phi^*_3}\;.
\end{eqnarray}
In contrast, the higher-order harmonics $v_4, v_5$ and $v_6$ arise from both the initial geometry and non-linear mixing of lower-order harmonics~\cite{Gardim:2011xv,Teaney:2012ke,Teaney:2013dta}. The relative contributions of linear and non-linear effects to these higher-order harmonics can be separated cleanly using the event-shape selection techniques. The details are discussed in Secs.~\ref{sec:33} and \ref{sec:34}.

The presence of large EbyE fluctuations of $\varepsilon_n$ and $\Phi_n^*$ also has consequence on the anisotropy of jet production at high $\pT$. Since the energy loss of a jet depends on the length and local energy density along its path traversing the medium, jet production rate is expected to be sensitive to fluctuations of the event shape in the initial state. Recently model calculations predicts sizable $v_1$--$v_6$ for jet production in A+A collisions at high $\pT$~\cite{Zhang:2012mi,Zhang:2012ie,Jia:2012ez}.

\section{Results}
\label{sec:3}
\subsection{Differential measurements of single flow harmonics $v_n$}
\label{sec:31}
Until recently, most flow studies were aimed at event-averaged $v_n$ coefficients, measured differentially as a function of $\pT$, $\eta$, centrality and particle species~\cite{Adare:2011tg,star:2013wf,Aamodt:2011by,ALICE:2011ab,Aad:2012bu,Chatrchyan:2012wg}. Many experimental methods have been developed in these measurements, including the event-plane method $v_n$\{EP\}~\cite{Ollitrault:1992bk}, two-particle correlation method $v_n$\{2PC\}~\cite{Poskanzer:1998yz}, scalar-product method $v_n$\{SP\}~\cite{Adler:2002pu}, multi-particle cumulant method $v_n$\{2\}, $v_n$\{4\}...~\cite{Borghini:2001vi,Bilandzic:2010jr}, and lee-yang-zero method $v_n$\{LYZ\}~\cite{Bhalerao:2003xf}. These methods are all based on multi-particle correlation concept, but they are constructed to have very different sensitivity to flow fluctuations and non-flow effects. In a nut-shell, the higher-order cumulant ($v_n$\{2$k$\} for $k>1$) and lee-yang-zero methods suppress both non-flow effects and flow fluctuations. The $v_n$\{2PC\}, $v_n$\{SP\} and $v_n$\{2\} are closely related to the RMS value of $p(v_n)$:
\begin{equation}
\label{eq:310}
v_n\{\mathrm{2PC}\}\approx v_n\{\mathrm{SP}\}\approx v_n\{\mathrm{2}\}\approx\sqrt{\langle v_n^2\rangle}\;,
\end{equation}
and they are sensitive to not only flow fluctuations but also non-flow effects, however the latter can be suppressed by requiring a rapidity gap between pair of particles. The most popular method, $v_n$\{EP\}, is known to introduce non-trivial biases in the presence of flow fluctuations~\cite{Alver:2008zza}, and hence should be used with caution. For all practical purpose, it can be replaced by $v_n$\{SP\}. More detailed discussion and comparison of these methods can be found in Ref.~\cite{Voloshin:2008dg,Luzum:2013yya}.

\begin{figure}
\includegraphics[width=1\linewidth]{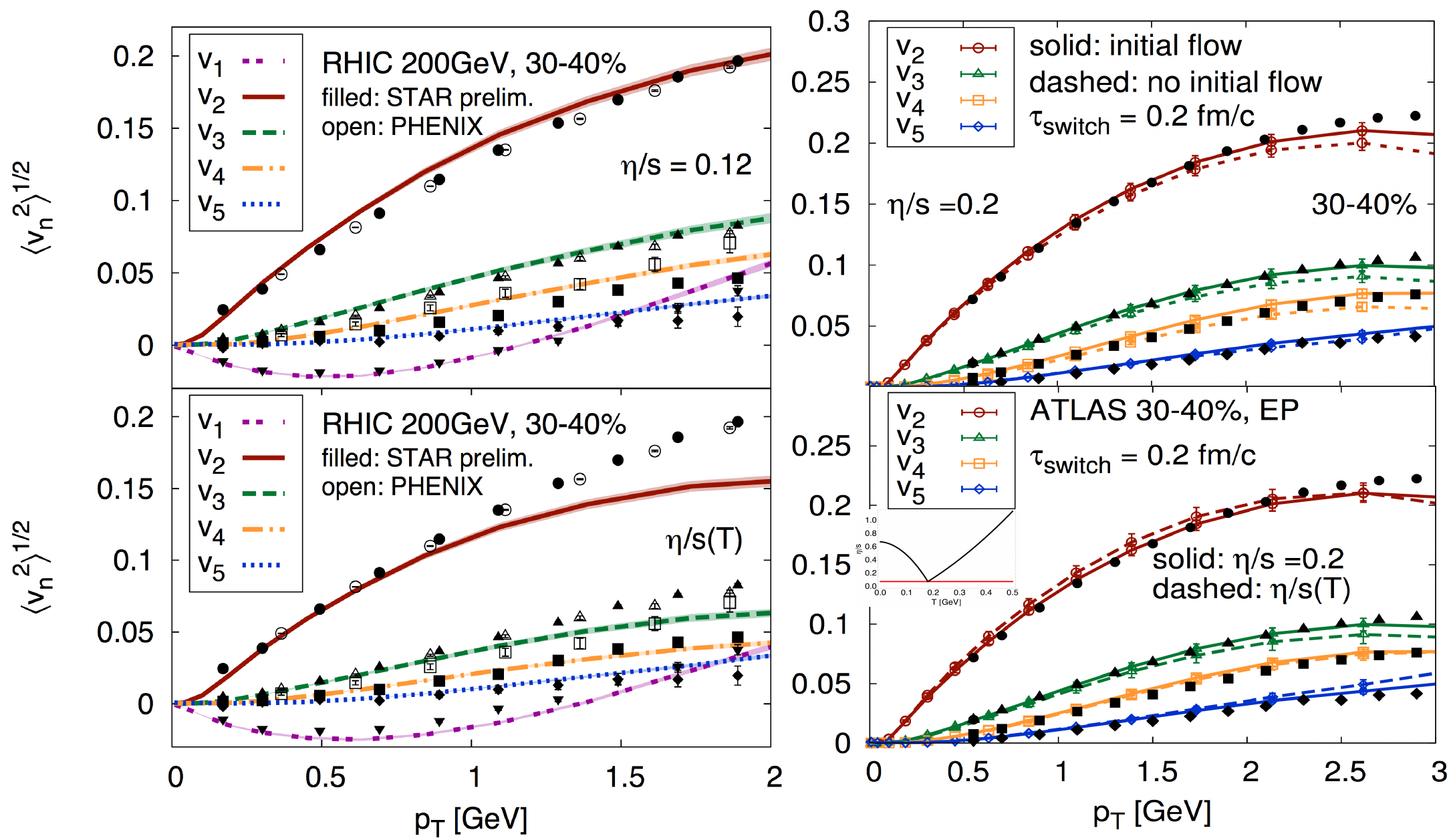}
\caption{\label{fig:res1} Calulation of $v_n(\pT)$ for $n=1$--5 from viscous hydro-calculation based on IP-Glamas initial condition~\cite{Gale:2012rq}, compared with data rom RHIC (left column) with constant $\eta/s$ (top-left) and temperature dependent $\eta/s$ (bottom-left) and from LHC (right column) with initial flow (top-right) and temperature dependent $\eta/s$ (bottom-right). The $v_1$ is the mean value, while $v_2$--$v_6$ are the RMS values.}
\end{figure}

Last few years (since 2010) also witnessed rapid development of hydrodynamic modeling of heavy ion collisions~\cite{Gale:2012rq}. Confronted with large amount of $v_n$ data, theorists are able to fine tune their models in terms of both the initial conditions and the hydrodynamic response. For example, when the first $v_3$ measurement was obtained by the PHENIX and ALICE collaboration~\cite{Adare:2011tg,ALICE:2011ab}, it became clear that the commonly used MC-KLN initial condition was unable to simultaneously describe the $v_2$ and $v_3$ data. As more detailed differential $v_n$ data became available, most early initial geometry models (prior to 2010) have been ruled out. The recently developed IP-Glasma model~\cite{Schenke:2012wb} takes into account gluons field fluctuations inside nucleons and the associated gluon saturation effects. As shown in Fig.~\ref{fig:res1}, the IP-Glasma initial condition combined with viscous hydrodynamic evolution, describe a large set of the measured $v_n(\pT,\mathrm{centrality})$ spectrum. However, this description is not perfect everywhere, in particular for dipolar flow $v_1$ at LHC (see~\cite{Gale:2012rq}, but not shown here) and higher-order harmonics. 

\begin{figure}
\includegraphics[width=1\linewidth]{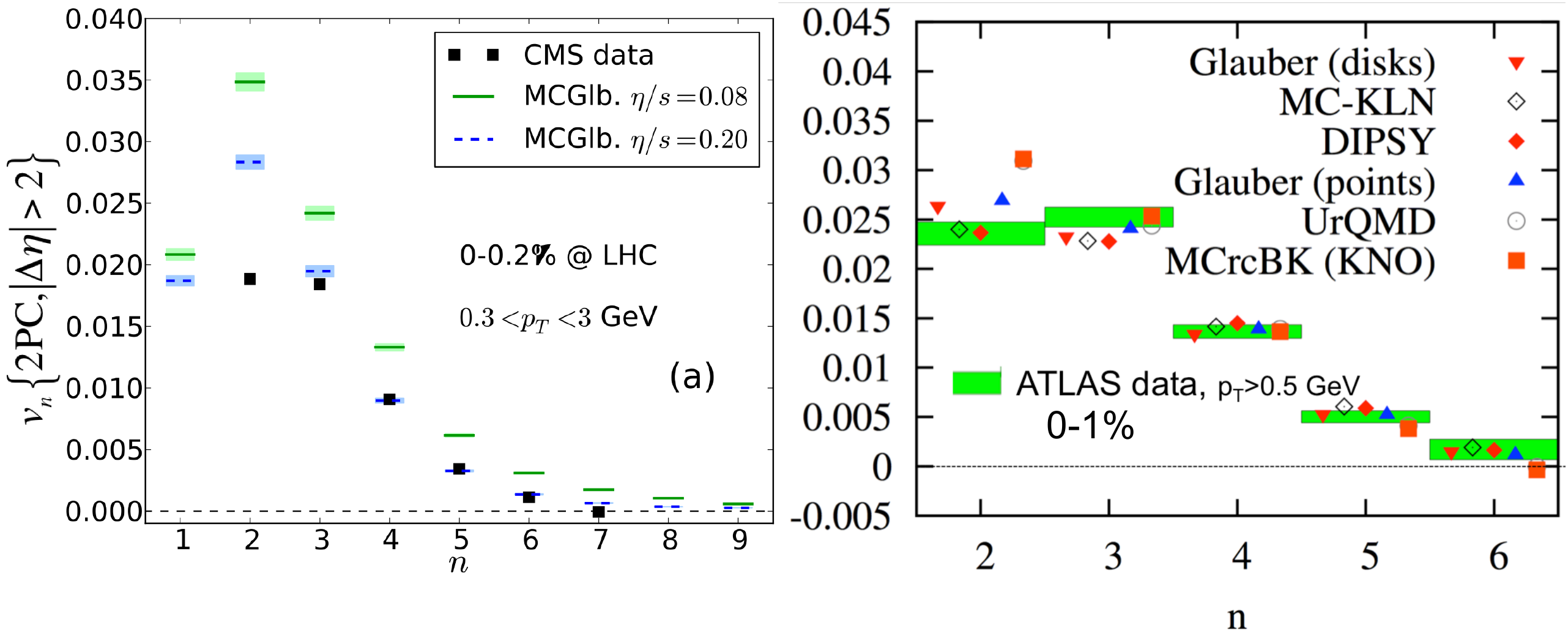}
\caption{\label{fig:res2} The $\pT$-integrated $v_n$ vs. $n$ in ultra-central Pb+Pb collisions from CMS for 0-0.2\% centrality (left) and ATLAS for 0-1\% centrality (right), compared with hydrodynamic model calculations with different initial conditions. Figures are taken from Ref.~\cite{CMS:2013bza} and Ref.~\cite{Luzum:2012wu}, respectively.}
\end{figure}

\begin{figure}
\includegraphics[width=1\linewidth]{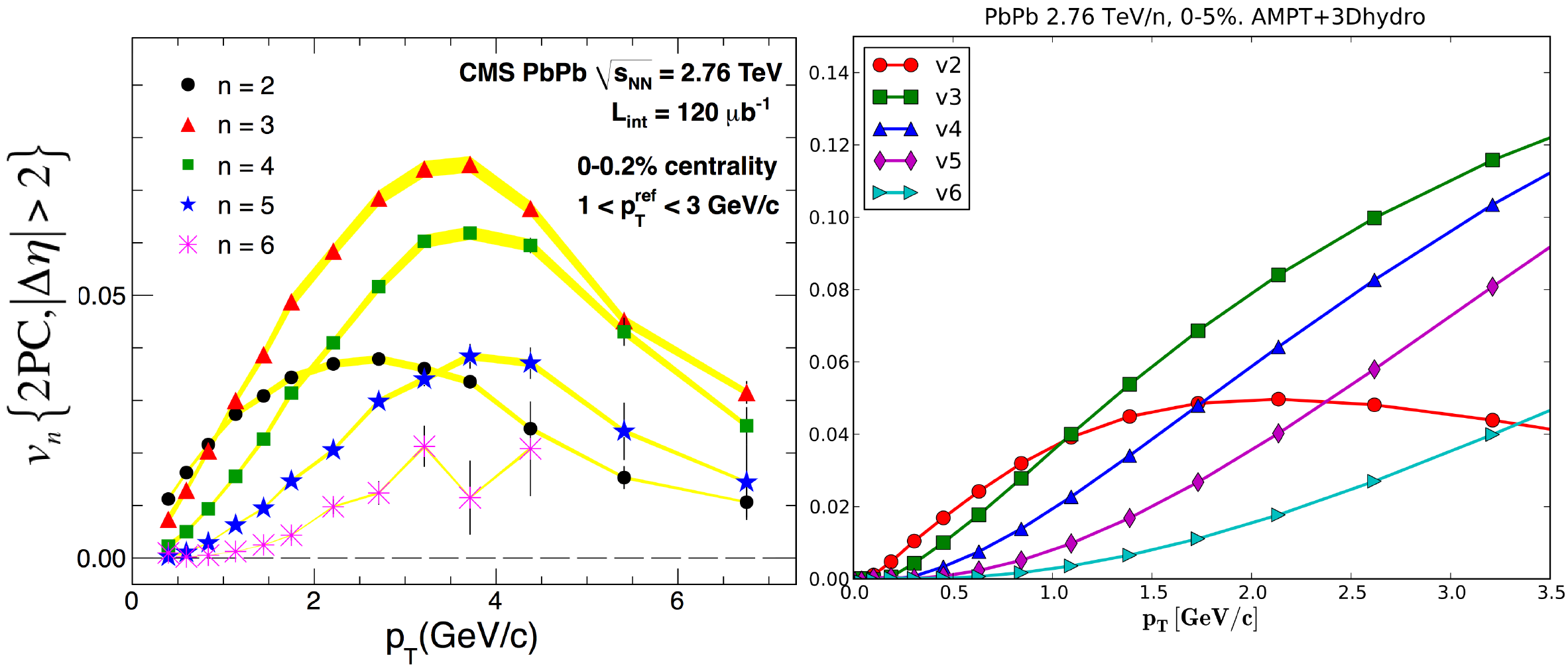}
\caption{\label{fig:res3} The $v_n(\pT)$ for $n=2$--6 in ultra-central Pb+Pb collisions from CMS for 0-0.2\% centrality~\cite{CMS:2013bza} and from 3+1D ideal hydrodynamic model calculations with initial conditions from AMPT model~\cite{pang,Pang:2012he}.}
\end{figure}

One challenge for the theory is the $v_n$ spectrum in ultra-central collisions, events in 0-1\% centrality or less~\cite{Aad:2012bu,CMS:2013bza}. The original motivation is that the initial conditions in these collisions are predominantly generated by fluctuations such that the magnitudes of first several $\varepsilon_n$ are comparable, and that the hydrodynamic response is expected to be linear $v_n\propto \varepsilon_n$ for all harmonics~\cite{Qiu:2011iv}~\footnote{The $\varepsilon_n$ of all order are comparable when calculated with $r^2$ weight, but increase gradually with $n$ for $r^n$ weight.}. Hence these collisions are expected to provide better constraints on the mechanism of the density fluctuations. The precision data from CMS and ATLAS show several features that are not described by models: 1) $v_3$ is comparable or larger than $v_2$ as shown in Fig.~\ref{fig:res2}. This is challenging since naively one expect $\varepsilon_3\approx \varepsilon_2$, but the $v_3$ should suffer larger viscous correction. 2) The $v_2(\pT)$ has very different shape comparing to other harmonics (Fig.~\ref{fig:res3}). It peaks at much lower $\pT$, around 1.5 GeV compare to 3-4 GeV for other harmonics. 3) The $v_2$ is also observed to break the factorization relation at 20-30\% level, while such breaking is much less in other centrality and for higher-order harmonics in all centrality~\cite{Aad:2012bu,CMS:2013bza}. The current hydro calculations describe the factorization data for $v_2$ but over-predict those for the $v_3$~\cite{Heinz:2013bua}. 

In a recent work, G. Denicol {\it et.al.} suggest that the nucleon-nucleon correlation and effects of bulk viscosity can reduce the $v_2$ values relative to $v_3$, thus partially but not completely resolving the hierarchy problem in Fig.~\ref{fig:res3}~\cite{Denicol:2014ywa}. It is possible that the longitudinal fluctuations and event-plane decorrelation effects might be responsible for the smallness of $v_2$ and the strange $v_2(\pT)$ shape, as indicated by a recent 3+1D hydro calculation that include longitudinal dynamics~\cite{pang,Pang:2012he}.

The $v_n$ data is precise enough to also improve the present modeling of other aspects of space-time evolution of the collisions, such as the early-time dynamics and initial flow, the temperature dependence of $\eta/s$ (see also Fig.~\ref{fig:res3}), and non-linear hydrodynamic response for $v_2$ or $v_3$~\cite{Gale:2012rq,Retinskaya:2013gca,Niemi:2014ita,harriqm2014}.

\subsection{Flow distribution $p(v_n)$}
\label{sec:32}
Due to large fluctuations in the initial density profile, $\varepsilon_n$ varies strongly event to event. Calculations based on a Monte-Carlo (MC) Glauber model show that, even for events in a very narrow centrality interval, $\varepsilon_n$ can fluctuate from zero to several times its mean value, leading to a very broad probability density distribution $p(\varepsilon_n)$~\cite{Aad:2013xma}. In central and mid-central collisions where the $\npart$ is large and flow response is approximately linear, the fluctuations of eccentricity and flow coefficients can be approximated by a 2-D Gaussian~\cite{Voloshin:2007pc}:
\begin{eqnarray}
\label{eq:pvn1}
p(\vec{\varepsilon}_n) \approx \frac{1}{2\pi\delta^2_{_{\varepsilon_n}}} e^{-\left(\vec{\varepsilon}_n-\vec{\varepsilon}_n^{{\;0}}\right)^2 \big{/}\left(2\delta^2_{_{\varepsilon_n}}\right)}\;,p(\vec{v}_n) \approx \frac{1}{2\pi\delta^2_{_{v_n}}} e^{-\left(\vec{v}_n-\vec{v}_n^{\;0}\right)^2 \big{/}\left(2\delta^2_{_{v_n}}\right)}\;,
\end{eqnarray}
where the $\vec{\varepsilon}_n^{\mathrm{\;0}}$ and $\vec{v}_n^{\mathrm{\;0}}$ represent the eccentricity and flow vector associated with average geometry in the reaction plane, and $\delta_{_{\varepsilon_n}}$ or $\delta_{_{v_n}}$ reflects the width of the fluctuations. Integration of this function over the azimuthal angle gives the one-dimensional (1D) probability density of $v_n=|\vec{v}_n|$ in the form of the Bessel--Gaussian (B-G) function~\cite{Voloshin:1994mz,Voloshin:2007pc}:
\begin{eqnarray}
\label{eq:pvn2}
p(v_n) =\frac{v_n}{\delta_{_{v_n}}^2}e^{-\frac{(v_n)^2+(v_n^{0})^2}{2\delta_{_{v_n}}^2}} I_0\left(\frac{v_n^{0}v_n}{\delta_{_{v_n}}^2}\right)\;,
\end{eqnarray}
where $I_0$ is the modified Bessel function of the first kind. If the $\beta\equiv v_n^{\mathrm{0}}/\delta_{_{v_n}}\ll1$ which is suitable for $p(v_2)$ in central collision or for $p(v_3)$, the $ v_n^{\mathrm{0}}$ can be absorbed into the width parameter~\cite{Aad:2013xma}:
\begin{eqnarray}
\label{eq:pvn2b}
p(v_n) =\frac{v_n}{\delta^{\prime 2}_{_{v_n}}} e^{-v^{\mathrm{\;2}}_n/\left(2\delta^{\prime 2}_{_{v_n}}\right)}+O\left(\beta v_n/\delta_{_{v_n}}\right)^4,
\delta^{\prime 2}_{_{v_n}} = \delta^{2}_{_{v_n}} {\textstyle \left(1-\beta^2/2\right)}^{-1}.
\end{eqnarray}
Thus when fluctuation is large, the value of $v_n^{\mathrm{0}}$ is constrained mainly by the tail of the distribution.

\begin{figure}
\includegraphics[width=0.52\linewidth]{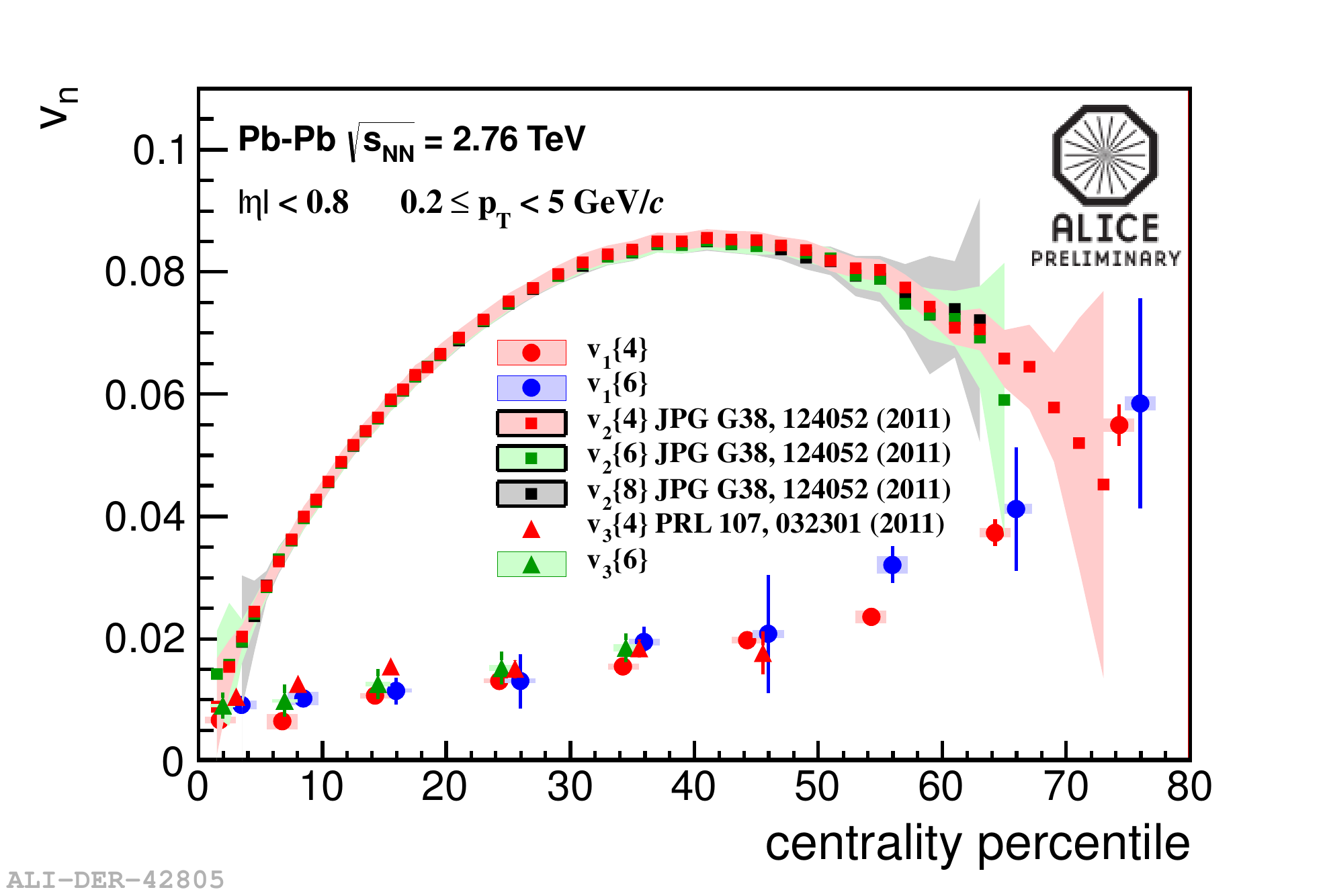}\includegraphics[width=0.52\linewidth]{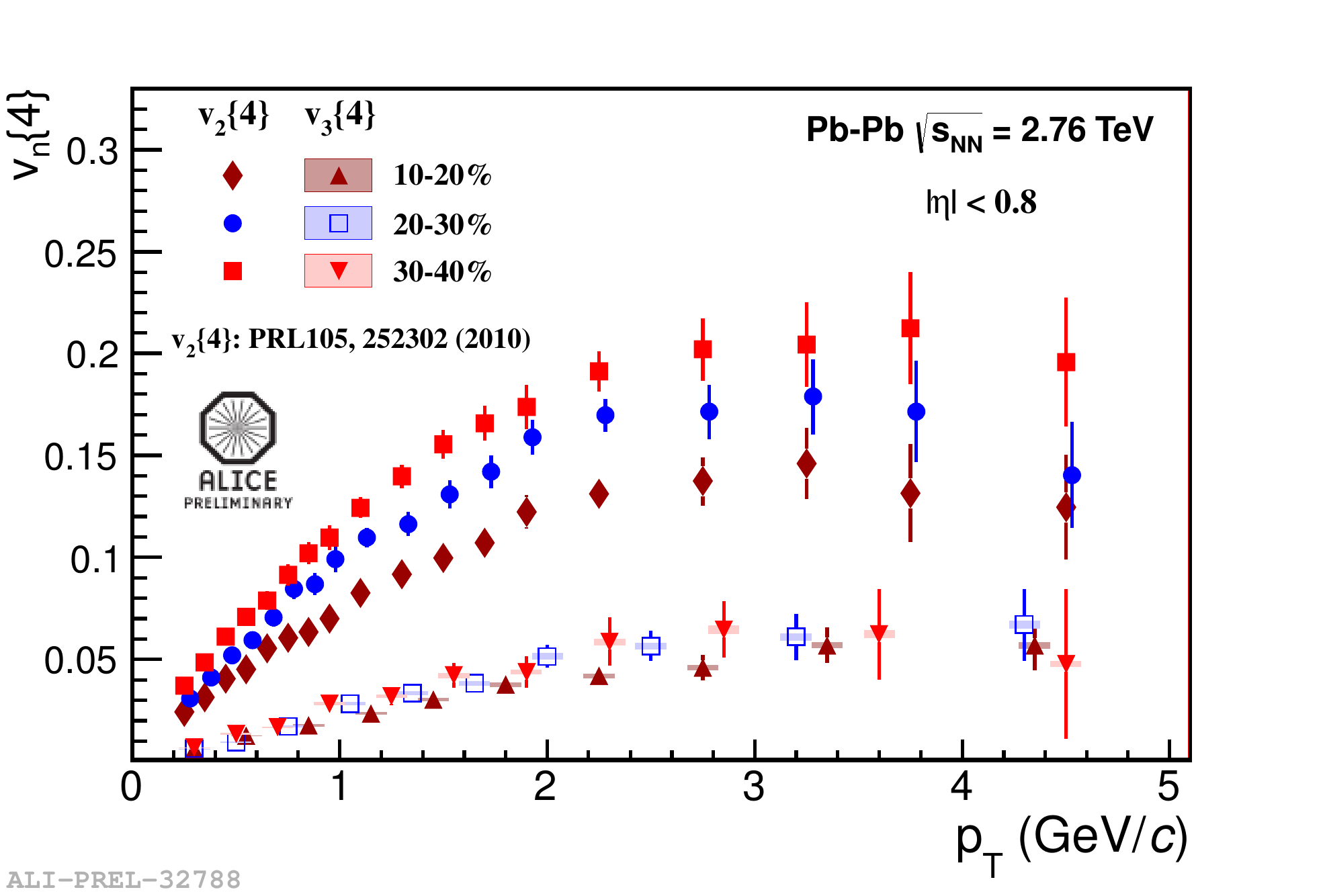}
\caption{\label{fig:res21} ALICE measurements of (left) centrality dependence of $v_1$, $v_2$ and $v_3$ estimated with multi-particle cumulants, and (right) transverse momentum dependence of $v_2$ and $v_3$ estimated with four-particle cumulants~\cite{Bilandzic:2012an}.}
\end{figure}

Traditionally, the information of $p(v_n)$ has been inferred from the multi-particle cumulants, $v_n$\{2\}, $v_n$\{4\} and so on. The first four of these cumulants can be expressed as~\cite{Voloshin:2007pc}:
\begin{eqnarray}
\nonumber
&&\hspace*{-1cm}v_n\{2\}^2\equiv\langle v_n^2\rangle\;,\\\nonumber
&&\hspace*{-1cm}v_n\{4\}^4\equiv-\langle v_n^4\rangle+2\langle v_n^2\rangle^2\;,\\\nonumber
&&\hspace*{-1cm}v_n\{6\}^6\equiv\left(\langle v_n^6\rangle-9\langle v_n^4\rangle\langle v_n^2\rangle+12\langle v_n^2\rangle^3\right)/4\;,\\\label{eq:pvn3}
&&\hspace*{-1cm}v_n\{8\}^8\equiv-\left(\langle v_n^8\rangle-16\langle v_n^6\rangle\langle v_n^2\rangle-18\langle v_n^4\rangle^2+144\langle v_n^4\rangle\langle v_n^2\rangle^2-144\langle v_n^2\rangle^4\right)/33\;.\;\;\;
\end{eqnarray}
The $\langle v_n^{2k}\rangle$ is calculated as the cosine average of azimuthal angle of all combination of $2k$ particles, and in the absence of non-flow, it is equivalent to the corresponding moment of the $v_n$ distribution:
\begin{eqnarray}
\label{eq:pvn3b}
\langle v_n^{2k}\rangle=\left\langle\cos (\sum_{j=1}^{k}n(\phi_{2j}-\phi_{2j+1}))\right\rangle\equiv\int v_n^{2k} p(v_n)dv_n,
\end{eqnarray}
where $\langle..\rangle$ denote the average over all combinations in a event then over all events. For B-G distribution, these cumulants have a particularly simple form~\cite{Voloshin:2007pc}:
\begin{equation}
\label{eq:pvn4}
v_n\{2k\}= \left\{\begin{array}{ll} \sqrt{\left(v_n^{\mathrm{0}}\right)^2+2\delta^2_{_{v_n}}} & k=1\\
v_n^{\mathrm{0}} & k>1\\   \end{array}\right.
\end{equation} 
Hence the observation that $v_n\{4\}\approx v_n\{6\}\approx v_n\{8\}$, see Figure~\ref{fig:res21} for $v_2$, has been used as evidence that flow fluctuation is Gaussian. In this case, assuming that the non-flow contribution is small, the $v_n$\{2\} and $v_n$\{4\} can be used to extract $v_n^{\mathrm{0}}$ (the component associated with average geometry) and $\delta_{_{v_n}}$ (component associated with fluctuation).

\begin{figure}
\center
\includegraphics[width=0.9\linewidth]{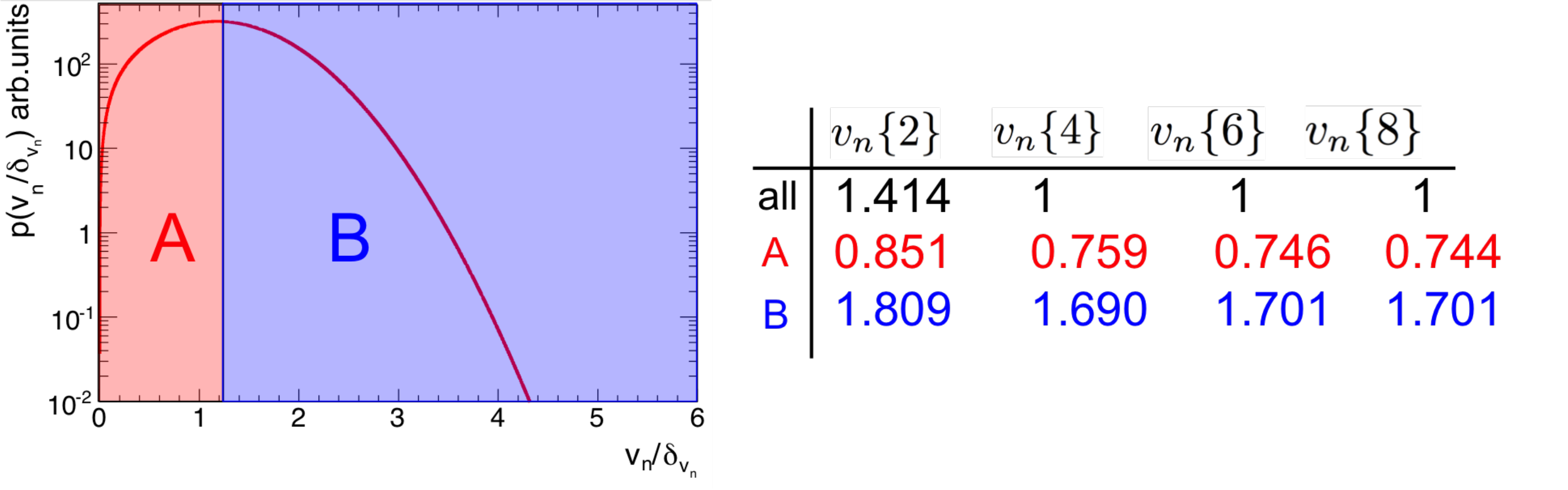}
\caption{\label{fig:bg} The distribution of Bessel-Gaussian function requiring $v_n^{0}=\delta_{_{v_n}}$. This distribution is divided into two halves with equal integral, labeled by A and B, respectively. The table at the right panel summarizes the values of cumulants $v_n\{2k\}$ for the full distribution and each half in units of $\delta_{_{v_n}}$, calculated via Eqs.~\ref{eq:pvn3}.}
\end{figure}

One limitation with the cumulant framework is that it is not possible to describe $v_n$ fluctuation using a small set of cumulants such that higher-order cumulants systematically vanishes, as in a taylor expansion. The higher-order cumulants are obtained after cancellation between several large numbers, so it can have sizable systematic uncertainties. Furthermore, if $v_n^{\mathrm{0}}$ is large, $v_n\{2k\}$ for $k>1$ may not be very sensitive to significant deviations from Bessel--Gaussian distribution. To illustrate this, one example B-G distribution shown in Fig.~\ref{fig:bg} is divided into two equal halves, and the $v_n\{2k\}$ are calculated analytically for each half using Eqs.~\ref{eq:pvn3}. Despite the fact that the truncated distribution in each half is non-Gaussian, the higher-order cumulants for $k>1$ are very close to each other. Hence the similarity between $v_n\{4\}$, $v_n\{6\}$ and $v_n\{8\}$ can not be used to conclude the flow fluctuation is Gaussian~\footnote{For some other functional forms, we find that $v_n\{2k\}$ for $k>1$ are not the same, and some terms in the right-hand side of Eq.~(23) may even be negative.}

For collisions with small $\npart$ (p+A or peripheral A+A collisions), the fluctuations of $\varepsilon_n$ and $v_n$ deviate strongly from Gaussian and maybe described by a power-law function~\cite{Yan:2013laa}:
\begin{equation}
\label{eq:pvn5}
p(v_n) = 2\alpha v_n(1-v_n^2)^{\alpha-1}
\end{equation} 
where $\alpha$ is proportional to $\npart$. In this case, the collisions have no average geometry, nevertheless the $v_n\{2k\}$ for $k>1$ are generally non-zero and approximately equal to each other. The small but non-zero $v_1$\{4\}, $v_1$\{6\} and $v_3$\{4\} (shown in Fig.~\ref{fig:res21}) and $v_4$\{4\}~\cite{Bilandzic:2012an,ATLAS2014-027} may be related to the non-Gaussianity of the $p(v_n)$ distribution.
\begin{figure}
\includegraphics[width=1\linewidth]{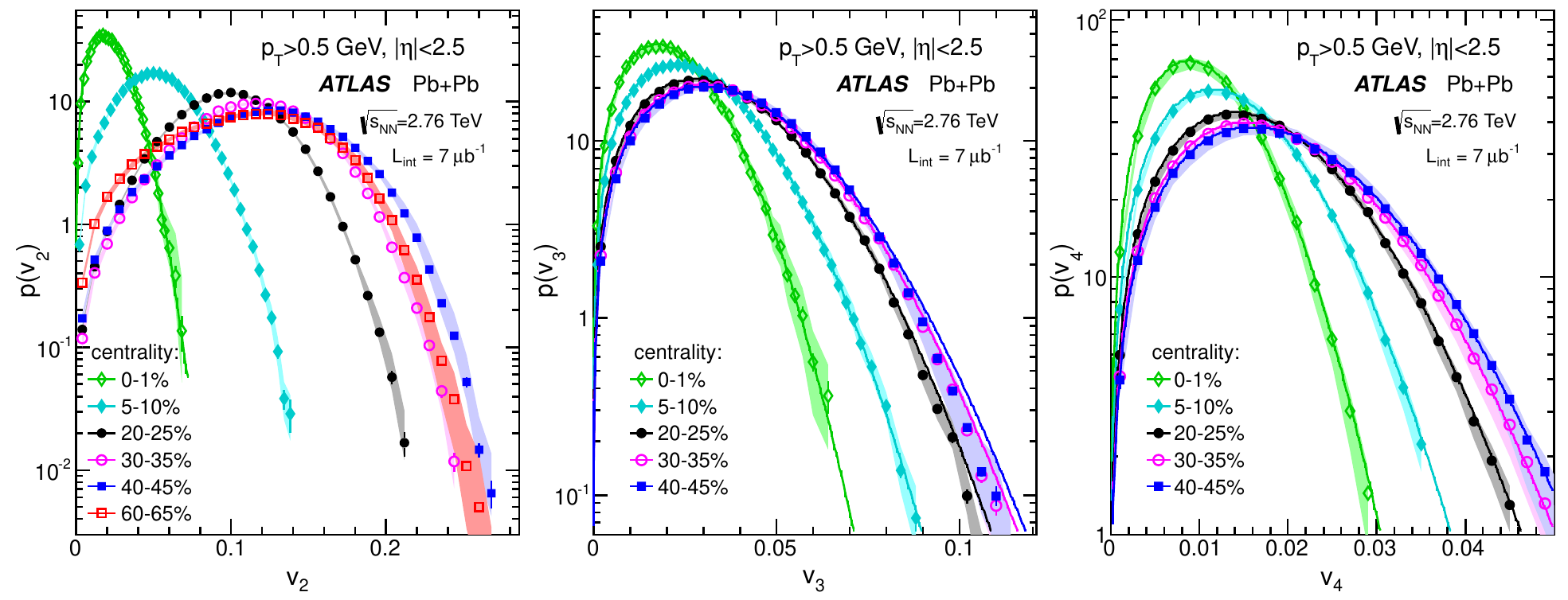}
\caption{\label{fig:res20} The event-by-event distributions of $v_2$, $v_3$ and $v_4$~\cite{Aad:2013xma}. Curves are fits to Bessel-Gaussian function Eq.~\ref{eq:pvn2} but with $v_n^{\;0}=0$ (or equivalently Eq.~\ref{eq:pvn2b}).}
\end{figure}

ATLAS employed a data-driven unfolding method to obtain directly the $p(v_n)$ distribution. In this method, an observed flow coefficient $\vec{v}_n^{\mathrm{\;obs}}$ is calculated for each event. The distribution $p(v_n^{\mathrm{obs}})$ contains the effects of smearing due to finite number of particles and non-flow. These effects are estimated by the response function $\vec{\delta}_n^{\;\mathrm{RF}}$, obtained from the difference of the $\vec{v}_n^{\mathrm{\;obs}}$ calculated separately from two sub-events at forward and backward pseudorapidities. These two quantities are related to the observed flow vector as:
\begin{eqnarray}
\nonumber
\vec{v}_n^{\mathrm{\;obs}} &=& (\vec{v}_n^{\mathrm{\;obs,F}}+\vec{v}_n^{\mathrm{\;obs,B}})/2\;,\\\label{eq:pvn5b}
\vec{\delta}_n^{\mathrm{\;RF}} &=& (\vec{v}_n^{\mathrm{\;obs,F}}-\vec{v}_n^{\mathrm{\;obs,B}})/2\;.
\end{eqnarray}
Since the statistical smearing and most non-flow effects are not correlated between the two subevents, the distribution of observed flow vector is simply the convolution of true flow vector and the response function: 
\begin{eqnarray}
\label{eq:pvn5c}
p(\vec{v}_n^{\mathrm{\;obs}})=p(\vec{v}_n)\otimes p(\vec{\delta}_n^{\mathrm{\;RF}}). 
\end{eqnarray}
Consequently, the truth flow distribution can be obtained by a standard unfolding procedure such a Bayesian unfolding. A detailed study based on HIJING and AMPT simulation~\cite{Lin:2004en} shows that $\vec{\delta}_n^{\mathrm{\;RF}}$ appears as a random Gaussian smearing of the underlying flow distribution, justifying the unfolding procedure.

\begin{figure}
\centering
\includegraphics[width=0.8\linewidth]{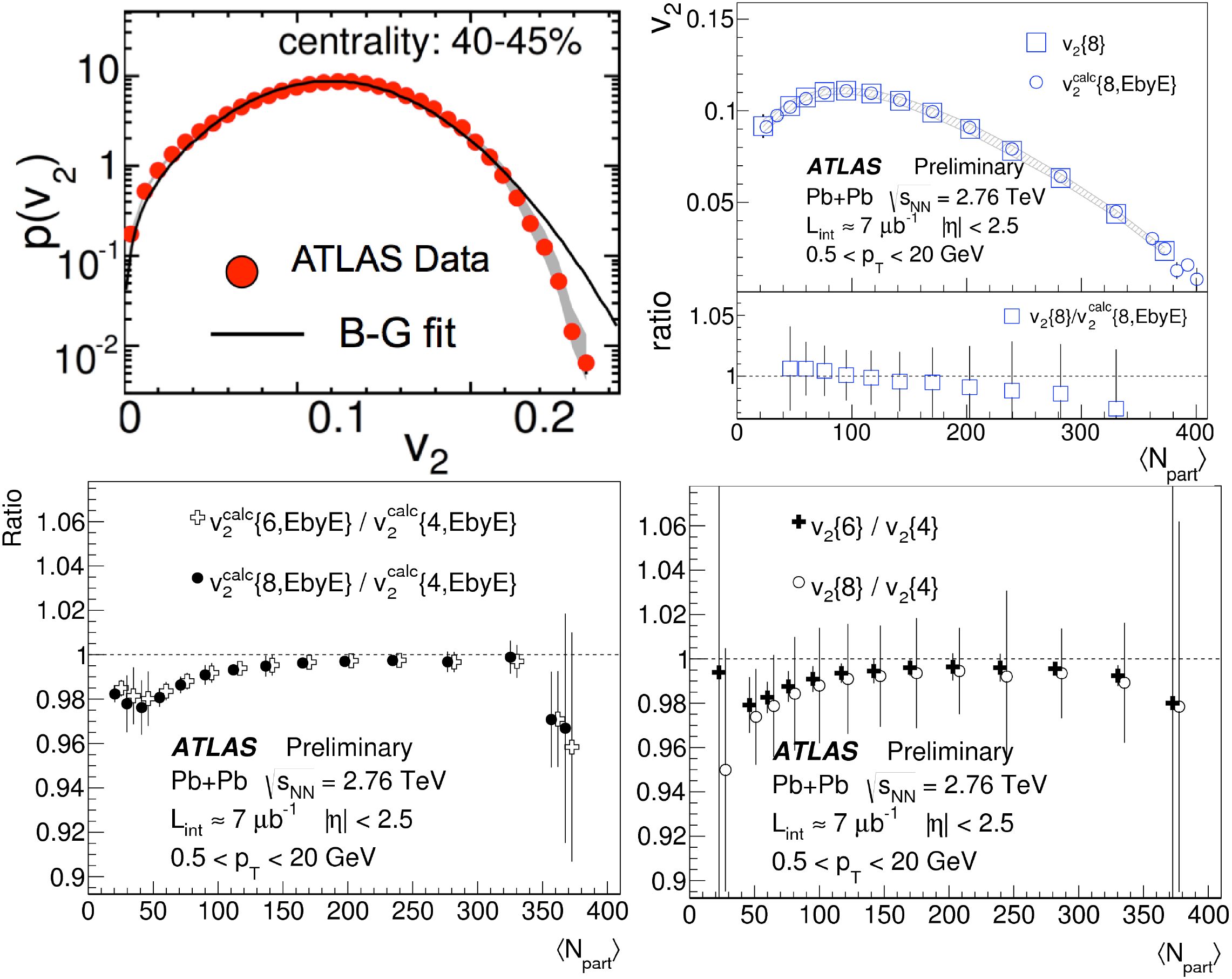}
\caption{\label{fig:res22} (Top-left) The $p(v_2)$ distribution and associated fit by Bessel-Gaussian function. (Top-right) Centrality dependence of $v_2$\{8\} obtained from multi-particle correlation method compared with that directly calculated from the $p(v_2)$. (Bottom) The ratios of the $v_2$\{6\}/$v_2$\{4\} and $v_2$\{8\}/$v_2$\{4\} from the two methods. Plots taken from Ref.~\cite{Aad:2013xma,ATLAS2014-027}.}
\end{figure}

Figure~\ref{fig:res20} shows the $p(v_2)$, $p(v_3)$ and $p(v_4)$ from ATLAS in several centrality intervals for Pb+Pb collisions. The $p(v_3)$ and $p(v_4)$ distributions are well described by the B-G functions. The $p(v_3)$ distributions suggest a small but non-zero $v_3$\{4\} as indicated from the deviation of the data from a pure gaussian function Eq.~\ref{eq:pvn2b}. The $p(v_2)$ distributions show significant deviations from B-G function in mid-central and peripheral collisions. One example is shown in the top-left panel of Fig.~\ref{fig:res22}. This deviation leads to a 2\% difference between $v_2$\{4\} and $v_2$\{6\}, but no difference between $v_2$\{6\} and $v_2$\{8\} (bottom panels of Fig.~\ref{fig:res22}), implying that when $v_n^{\mathrm{0}}$ is large, $v_n\{2k\}$ for $k>1$ responds slowly to deviation from B--G function. The small change of $v_n\{2k\}$ can be easily buried under the experimental systematic uncertainties. 

Figure~\ref{fig:res22} also compares the $v_2\{2k\}$ obtained from $2k$ particle correlations (left part of Eq.~\ref{eq:pvn3b}) with that calculated directly from the $p(v_2)$ distribution (second part of Eq.~\ref{eq:pvn3b}). Excellent agreement within 1\% is observed across the full centrality range. Hence cumulants calculated from $p(v_n)$ distribution provides equivalent information as the traditional method, but it is more intuitive and transparent in the estimation of systematic uncertainties.

\begin{figure}
\centering
\includegraphics[width=1\linewidth]{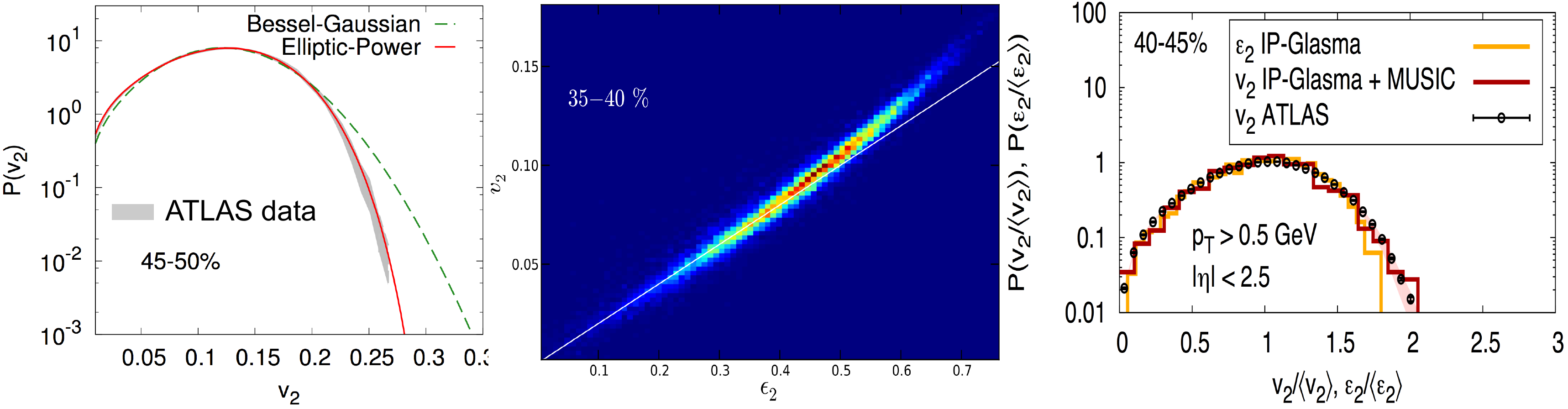}
\caption{\label{fig:res23} (left) The $p(v_2)$ data fit by Bessel-Gaussian and elliptic power function~\cite{yanliqm2014}. (Middle) $v_2$ vs. $\varepsilon_2$ from viscous hydro calculation with EbyE fluctuating initial condition~\cite{harriqm2014}. (Right) The scaled $v_2$ distribution compared to viscous hydro calculation with EbyE IP-Glasma initial conditions~\cite{Gale:2012rq,Gale:2012in}.}
\end{figure}

One reason for the non-Gaussian behavior is that the eccentricity distribution $p(\varepsilon_2)$ is bounded from above by $\varepsilon_2<1$. As a result, $p(\varepsilon_2)$ itself deviates strongly from B--G distribution and can instead be parameterized by a ``elliptic-power'' function~\cite{Yan:2014afa}. Since $v_2\propto\varepsilon_2$, the $p(v_2)$ distribution is also expected to fall faster than B--G function at large $v_2$ value (see top-left panel of Fig.~\ref{fig:res22}). But the problem is that the shape of the $p(\varepsilon_2)$ distributions from various initial geometry models can not be tuned to agree with $p(v_2)$ distribution across the full centrality range. In peripheral collisions, for example, the $p(\varepsilon_2)$ always falls faster than $p(v_2)$ in the tail of the distributions. There are two possible explanations: 1) Current modeling of $p(\varepsilon_2)$ is wrong, but response coefficient $k_2= v_2/\varepsilon_2$ is constant across the full $\varepsilon_2$ range~\cite{yanliqm2014}. In this case, the eccentricity distribution can be obtained as $p(\varepsilon_2)=p(v_2)/k_2$ (see left panel of Fig.~\ref{fig:res23}). 2) The modeling of $p(\varepsilon_2)$ is correct, but $k_2$ is a function of $\varepsilon_2$. A recent hydrodynamic model calculation~\cite{harriqm2014} suggests that the $k_2$ increases slightly at large $\varepsilon_2$ as shown in the middle panel of Fig.~\ref{fig:res23}, hence the $p(v_2)$ distribution is expect to decrease more slowly than the $p(\varepsilon_2)$ distribution, consistent with experimental observation~\cite{Aad:2013xma}. Similar non-linear behavior of $k_2$ is also observed in hydro calculations based on the IP-Glasma initial condition as shown in the right panel of Fig.~\ref{fig:res23}.

In most hydro calculation with fluctuating initial condition, a sizable spread has been observed in the correlation between EbyE $\varepsilon_n$ and EbyE $v_n$ as shown in Fig.~\ref{fig:res23}. This spread can be partially related to the fluctuations of event-plane angle as a function of $\pT$ and $\eta$: $\Phi_n(\pT,\eta)$, observed in hydrodynamic simulations~\cite{Gardim:2012im,Heinz:2013bua}. The correlation is often quantified by a linear correlation coefficient:
 \begin{eqnarray}
c(v_n,\varepsilon_n) = \frac{\left\langle (v_n-\langle v_n\rangle)(\varepsilon_n-\langle \varepsilon_n\rangle)\right\rangle}{\sqrt{\left\langle (v_n-\langle v_n\rangle)^2\right\rangle\left\langle (\varepsilon_n-\langle \varepsilon_n\rangle)^2\right\rangle}}
\end{eqnarray}
$c(v_n,\varepsilon_n)=1$ or -1 indicates perfect correlation or anti-correlation. Given that $\varepsilon_n$ can be defined with different radial weights, and that the correlation between these definitions also have significant spread, it is not surprising that the $v_n$--$\varepsilon_n$ correlation should not be perfect. However, it is not clear how much of this spread is due to the higher-order radial modes in the initial geometry or is generated dynamically in the final state (e.g. hydrodynamic noise or hadronic freezeout). This remains an open issue to be resolved in the future.

\begin{figure}
\centering
\includegraphics[width=1\linewidth]{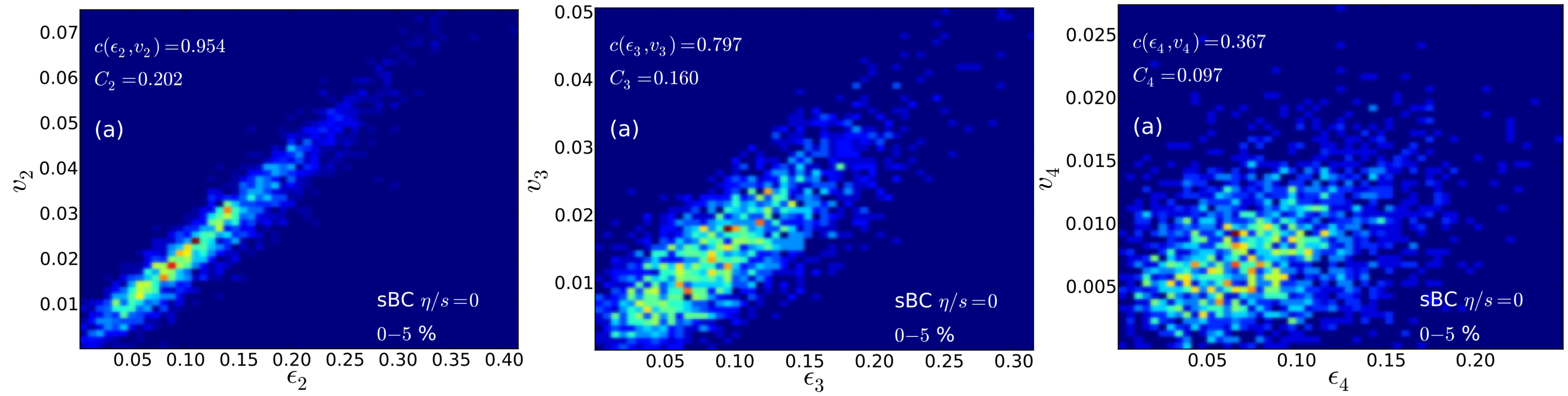}
\caption{\label{fig:res24} The pion $v_n$ vs. $\varepsilon_n$ for $n=2$, 3 and 4 in 0-5\% central Pb+Pb collisions from EbyE viscous hydro calculations~\cite{Niemi:2012aj}.}
\end{figure}

\subsection{Event-plane correlations $p(\Phi_n,\Phi_m,...)$}
\label{sec:33}
\begin{figure}
\centering
\includegraphics[width=1\linewidth]{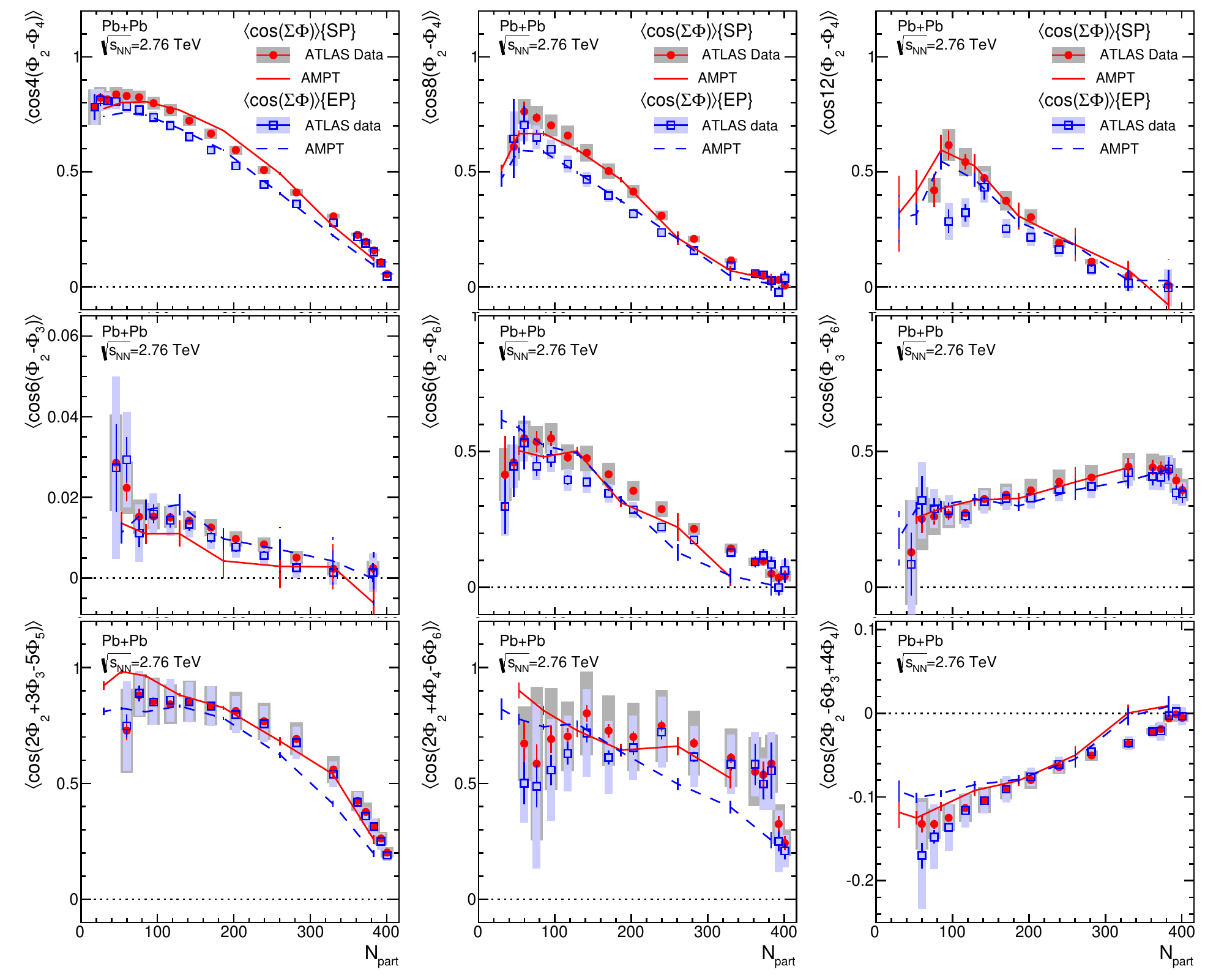}
\caption{\label{fig:res31} The event-plane correlations calculated from EP method (open symbols) and SP method (filled symbol)~\cite{Aad:2014fla} and compared with those obtained from the AMPT models~\cite{Bhalerao:2013ina}.}
\end{figure}

The correlation between different harmonic planes can in general be captured by multi-particle correlation as given by Eq.~\ref{eq:223}. A few such correlators have been measured by ALICE~\cite{Bilandzic:2012an}. In practice, it is often desirable to cast these correlators into different forms, such as the event-plane method and the scalar-product method used by ATLAS. Ignoring the statistical smearing and non-flow for the sake of simplicity, the correlators from these two methods reduce to:
\small{\begin{eqnarray}
\label{eq:ep1}
\hspace*{-2cm}\cos(\Sigma_k kc_k\Phi_k)\{\mathrm{EP}\} &\rightarrow& \left\langle \cos(c_1\Phi_1+2c_2\Phi_2...+lc_l\Phi_l)\right\rangle\\\nonumber\\\nonumber
\hspace*{-2cm}\cos(\Sigma_k kc_k\Phi_k)\{\mathrm{SP}\} &\rightarrow& \frac{\langle  v_1^{|c_1|} v_2^{|c_2|}...v_l^{|c_l|}\cos(c_1\Phi_1+2c_2\Phi_2...+lc_l\Phi_l)\rangle}{\sqrt{\langle v_1^{|2c_1|}\rangle\langle v_2^{|2c_2|}\rangle...\langle v_l^{|2c_l|}\rangle}}\\\nonumber
\hspace*{-2cm}&=&\frac{\langle v_1^{|c_1|} v_2^{|c_2|}...v_l^{|c_l|}\cos(c_1\Phi_1+2c_2\Phi_2...+lc_l\Phi_l)\rangle}{\langle v_1^{|c_1|} v_2^{|c_2|}...v_l^{|c_l|}\rangle}\frac{\langle v_1^{|c_1|} v_2^{|c_2|}...v_l^{|c_l|}\rangle}{\sqrt{\langle v_1^{|2c_1|}\rangle\langle v_2^{|2c_2|}\rangle...\langle v_l^{|2c_l|}\rangle}}\\\label{eq:ep1c}
\hspace*{-2cm}&\equiv&\cos(\Sigma_k kc_k\Phi_k)\{\mathrm{EP}\}_w \;\;\;\; F(v_1^{|c_1|},v_2^{|c_2|},...,v_l^{|c_l|})
\end{eqnarray}}\normalsize
Eq.~\ref{eq:ep1c} has been separated into two parts. The first part is the cosine average of the normalized distribution of $c_1\Phi_1+2c_2\Phi_2...+lc_l\Phi_l$, which is similar to Eq.~\ref{eq:ep1} except for the $v_n$ weights. The second part, denoted as $F(v_1^{|c_1|},v_2^{|c_2|},...,v_l^{|c_l|})$, is a factor that is always smaller than one due to Cauchy-Schwarz inequality. If the angular correlation is independent of the magnitude of the flow, the first part of Eq.~\ref{eq:ep1c} should equal Eq.~\ref{eq:ep1}. In this case, the correlator from the SP method should always be smaller than that given by the EP method. Experimental results as shown in Fig.~\ref{fig:res31}, on the other hand, suggest the opposite. This can happen only if the events with larger flow also have stronger angular correlations in order to compensate for the $F$ factor.

\begin{figure}
\centering
\includegraphics[width=1\linewidth]{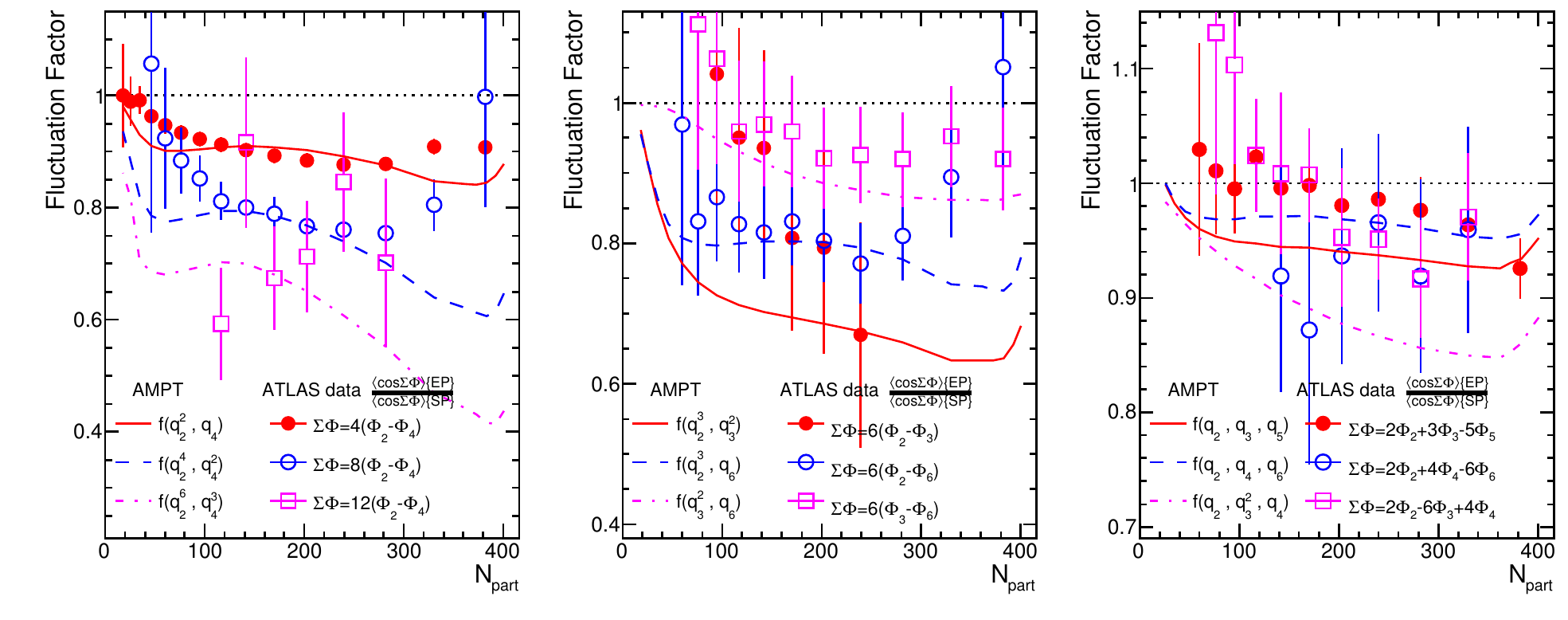}
\caption{\label{fig:res32} The $F$ factor in Eq.~\ref{eq:ep1c} from AMPT (lines) and ratios of the correlators between EP method and SP method from ATLAS (symbols) for the nine correlators shown in Fig.~\ref{fig:res31}.}
\end{figure}

To verify this, we estimate the value of $F$ in the AMPT model~\cite{Lin:2004en}, which was found to quantitatively describe the measured the EP correlators. The resulting factors, as shown in Fig.~\ref{fig:res32}, are always below one as expected. Interestingly, these factors are also found to be approximately the same as the ratio of the correlators between the two methods:
\begin{eqnarray}
\label{eq:ep2}
\frac{\cos(\Sigma_k kc_k\Phi_k)\{\mathrm{EP}\}}{\cos(\Sigma_k kc_k\Phi_k)\{\mathrm{SP}\}} \simeq F(v_1^{|c_1|},v_2^{|c_2|},...,v_l^{|c_l|})
\end{eqnarray}
From this, we obtain the following approximate empirical relation:
\begin{eqnarray}
\label{eq:ep2b}
\frac{\cos(\Sigma_k kc_k\Phi_k)\{\mathrm{EP}\}}{\cos(\Sigma_k kc_k\Phi_k)\{\mathrm{EP}\}_w} \simeq F^2(v_1^{|c_1|},v_2^{|c_2|},...,v_l^{|c_l|})
\end{eqnarray}
This relation suggests that the unweighted EP correlators are approximately a factor of $F^2$ weaker than the $v_n$-weighted case, but about half of this difference is cancelled out in the SP method (Eq.~\ref{eq:ep1c}). In the future, it would be useful to calculate $\cos(\Sigma_k kc_k\Phi_k)\{\mathrm{EP}\}_w$ and $F$ factor separately, in order to separate the EbyE fluctuation of angular correlations and EbyE fluctuation of $v_n$ magnitudes.

Several theory groups recently calculated the centrality dependence of EP correlators using hydrodynamic models~\cite{Gardim:2011xv,Teaney:2012ke,Qiu:2012uy,Teaney:2012gu,Teaney:2013dta}. The results of these calculations are in qualitative agreement with the experimental data. The dynamical origin of these correlators has been explained using the so-called single-shot hydrodynamics~\cite{Teaney:2012gu,Teaney:2012ke,Teaney:2013dta}, where small fluctuations are imposed on a smooth average geometry profile, and the hydrodynamic response to these small fluctuations is then derived analytically using a cumulant expansion method. In this analytical approach, the $v_4$ signal comprises a term proportional to the $\varepsilon_4$ (linear response term) and a leading non-linear term that is proportional to $\varepsilon_2^2$~\cite{Gardim:2011xv,Teaney:2012gu}:
\begin{eqnarray}
\label{eq:ep3a}
\hspace*{-1.5cm}v_4e^{i4\Phi_4} = a_{0}\; \varepsilon_4e^{i4\Phi^*_4}  + a_{1}\; \left(\varepsilon_2e^{i2\Phi_2^*}\right)^{2} +... =  c_0\;  e^{i4\Phi^*_4}  + c_{1}v_2^2e^{i4\Phi_2} +...\;,
\end{eqnarray}
where the second line of the equation is derived from Eq.~\ref{eq:224}, and $c_0=a_{0}\varepsilon_4$ denotes the linear component of $v_4$ and coefficients $a_0$, $a_1$ and $c_1$ are weak functions of centrality. This decomposition of the $v_4$ signal explains the measured correlation between $\Phi_2$ and $\Phi_4$, e.g. $\langle\cos 4k(\Phi_2-\Phi_4)\rangle$ ($k=1,2$, and 3). In the same manner, the $v_5$ signal comprises a linear component proportional to $\varepsilon_5$ and a leading non-linear term involving $v_2$ and $v_3$:
\begin{eqnarray}
\label{eq:ep3b}
\hspace*{-1.5cm}v_5e^{i5\Phi_5} = a_0 \varepsilon_5e^{i5\Phi^*_5} + a_1 \varepsilon_2e^{i2\Phi^*_2}\varepsilon_3e^{i3\Phi^*_3}+...= c_0  e^{i5\Phi^*_5} + c_1 v_2v_3e^{i(2\Phi_2+3\Phi_3)}+...\;
\end{eqnarray}
which explains the centrality dependence of $\langle\cos (2\Phi_2+3\Phi_3-5\Phi_5)\rangle$. Similarly, the $v_6$ signal is given by the following decomposition:
\begin{eqnarray}
\nonumber
v_6e^{i6\Phi_6} &=& a_0 \varepsilon_6e^{i6\Phi^*_6} + a_1 \left(\varepsilon_2e^{i2\Phi^*_2}\right)^3+a_2 \left(\varepsilon_3e^{i3\Phi^*_3}\right)^2+
a_3\varepsilon_2e^{i2\Phi^*_2}\varepsilon_4e^{i4\Phi^*_4}+...\\\label{eq:ep3c}
              &=& c_0  e^{i6\Phi^*_6} + c_1 v_2^3e^{i6\Phi_2}+c_2 v_3^2e^{i6\Phi_3}+c_3 v_2e^{i(2\Phi_2+4\Phi^*_4)}+...\;
\end{eqnarray}
This decomposition of the $v_6$ signal explains the measured correlations between $\Phi_6$ and $\Phi_2, \Phi_3$ and/or $\Phi_4$, such as $\langle\cos 6(\Phi_2-\Phi_6)\rangle$, $\langle\cos 6(\Phi_3-\Phi_6)\rangle$ and $\langle\cos(2\Phi_2+4\Phi_4-\Phi_6)\rangle$~\cite{Aad:2014fla}. In particular, the large signal of $\langle\cos(2\Phi_2+4\Phi_4-\Phi_6)\rangle$ observed in central collisions (0-5\% range) must arise from the non-linear coupling between $\Phi^*_2$ and $\Phi^*_4$ corresponding to the third term, since both $\langle\cos 4(\Phi_2-\Phi_4)\rangle$ and $\langle\cos 6(\Phi_2-\Phi_6)\rangle$ are very small in central collisions as shown in Fig.~\ref{fig:res31}. One exception is $\langle\cos(2\Phi_2-6\Phi_3+4\Phi_4)\rangle$, which can not be explained by a simple decomposition like Eqs.~\ref{eq:ep3a}--\ref{eq:ep3c}. This probably suggests other non-linear terms, possibly involving dipolar flow $v_1$, need to be included simultaneously. 
\subsection{Flow amplitude correlations $p(v_n,v_m)$}
\label{sec:34}
\begin{figure}
\centering
\includegraphics[width=1\linewidth]{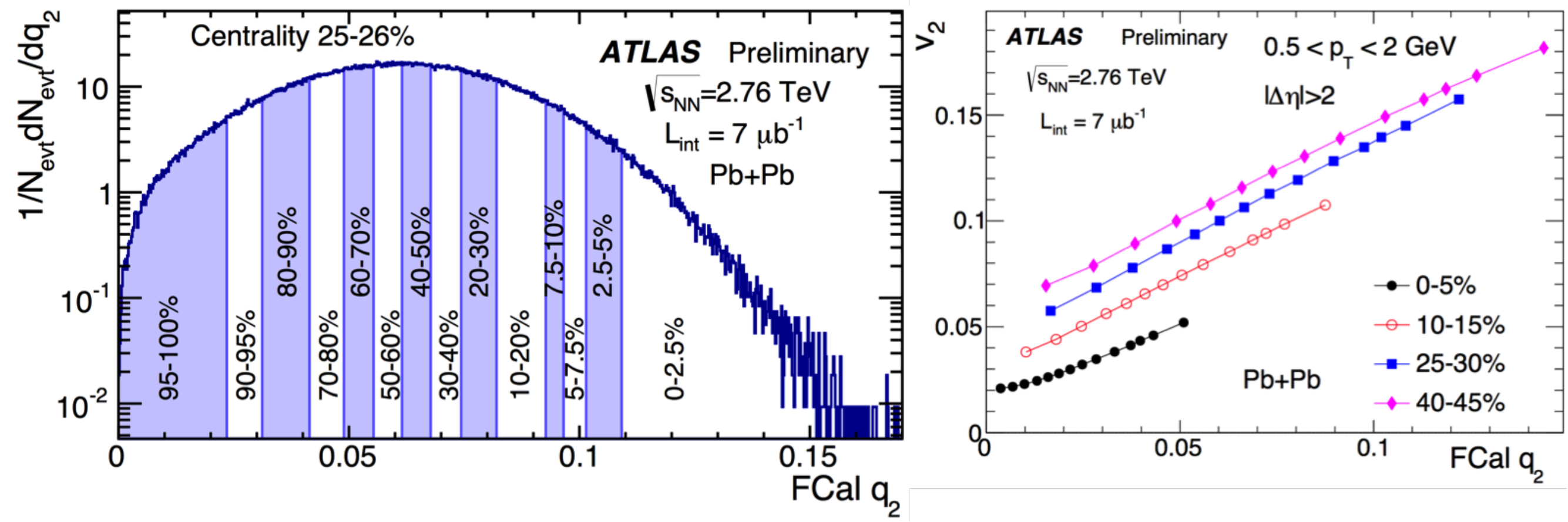}
\caption{\label{fig:res41} Left panel: The distribution of $q_2$ (or $v_2^{\mathrm{obs}}$) calculated in forward calorimeter ($3.3<|\eta|<4.9$) for events within a narrow fixed centrality range. This distribution is divided into different $q_2$ ranges. Right panel: Correlation of the mid-rapidity $v_2$ with $q_2$ for four centrality intervals, where $v_2$ is calculated from 2PC method for particles with $|\eta|<2.5$ and $|\Deta|>2$. Results taken from Ref.~\cite{ATLAS2014-022}.}
\end{figure}

The event-shape selection method~\cite{Schukraft:2012ah,Huo:2013qma} provides a more direct way of studying the correlation between $v_n$ and $v_m$. In one implementation of such method by the ATLAS experiment~\cite{ATLAS2014-022}, the azimuthal angle distribution of the transverse energy $\eT$ in the forward calorimeter over $3.3<|\eta|<4.9$ is expanded into a Fourier series for each event:
\begin{eqnarray}
\label{eq:res41}
2\pi\frac{d\eT}{d\phi} = （\Sigma \eT）\left(1+2\Sigma_{n=1}^{\infty}q_n\cos n(\phi-\Psi_n)\right)
\end{eqnarray}
where the reduced flow vector $q_n$ represents the $\eT$-weighted raw flow coefficients $v_n^{\mathrm{obs}}$, $q_n=\Sigma \left(\eT v_n^{\mathrm{obs}}\right)/\Sigma \eT$. Traditionally, events are divided according to $\Sigma \eT$ into different centrality classes. But here events are further divided according to the second harmonic modulation of the $\eT$ distribution, $q_2$. This classification separates events with similar multiplicity but with very different ellipticity. The values of $v_n$ are then calculated with charged particles at mid-rapidity ($|\eta|<2.5$) using a two-particle correlation method. The non-flow effects are suppressed by a rapidity gap $|\Deta|>2$ between charged particle pairs. Figure~\ref{fig:res41} shows the performance of event-shape selection on $q_2$ in the ATLAS detector. The values of $v_2$ are varied by up to a factor of three for events with similar centrality. 

\begin{figure}
\centering
\includegraphics[width=1\linewidth]{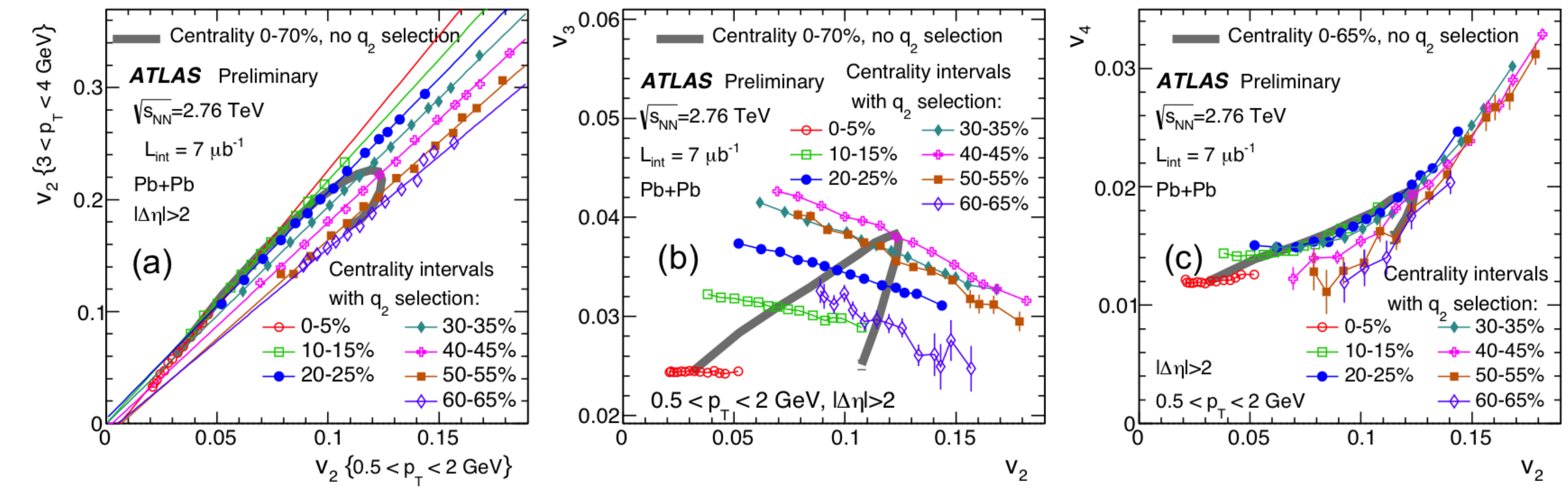}
\caption{\label{fig:res42} The correlation of (panel a) $v_2$ in two different $\pT$ ranges, (panel b) $v_3$ and $v_2$ in the same $\pT$ range and (panel c) $v_4$ and $v_2$ in the same $\pT$ range. The data points in each centrality interval correspond to the fourteen event classes with different ellipticity selected via an event-shape engineering technique. These data are overlaid with the centrality dependence without event-shape selection (think grey lines). The thin solid straight lines in the left panel represent a linear fit of the data in each centrality, and error bars represent the statistical uncertainties. Results taken from Ref.~\cite{ATLAS2014-022}.}
\end{figure}

The correlation of $v_2$ between two different $\pT$ ranges, presented in Fig.~\ref{fig:res42}(a), shows a boomerang-like centrality dependence, reflecting a stronger viscous correction for the centrality dependence of $v_2$ at higher $\pT$. In contrast, the same correlation within a narrow centrality interval is found to be always linear. This linearity indicates that viscous effects are controlled by the system size not its overall shape. Hence the event-shape selection method provides a more precise control on the viscosity effects.

The $v_3$--$v_2$ correlations, shown in Fig.~\ref{fig:res42}(b), reveal a surprising anti-correlation between the triangularity and ellipticity of the initial geometry. This anti-correlation is very similar in magnitude to the correlation between $\varepsilon_3$ and $\varepsilon_2$, suggesting that this correlation reflects mostly initial geometry effects, which is expected from the dominance of linear hydrodynamic response for $v_2$ and $v_3$. The geometric origin of this anti-correlation is easy to understand. As the overlap region becomes more elliptic or elongated, the fluctuation of triangularity is constrained: instead of fitting one large triangular for a circular overlap region (small $\varepsilon_2$), one can only fit multiple, uncorrelated small triangles for an elliptic geometry (large $\varepsilon_2$).

\begin{figure}
\centering
\includegraphics[width=1\linewidth]{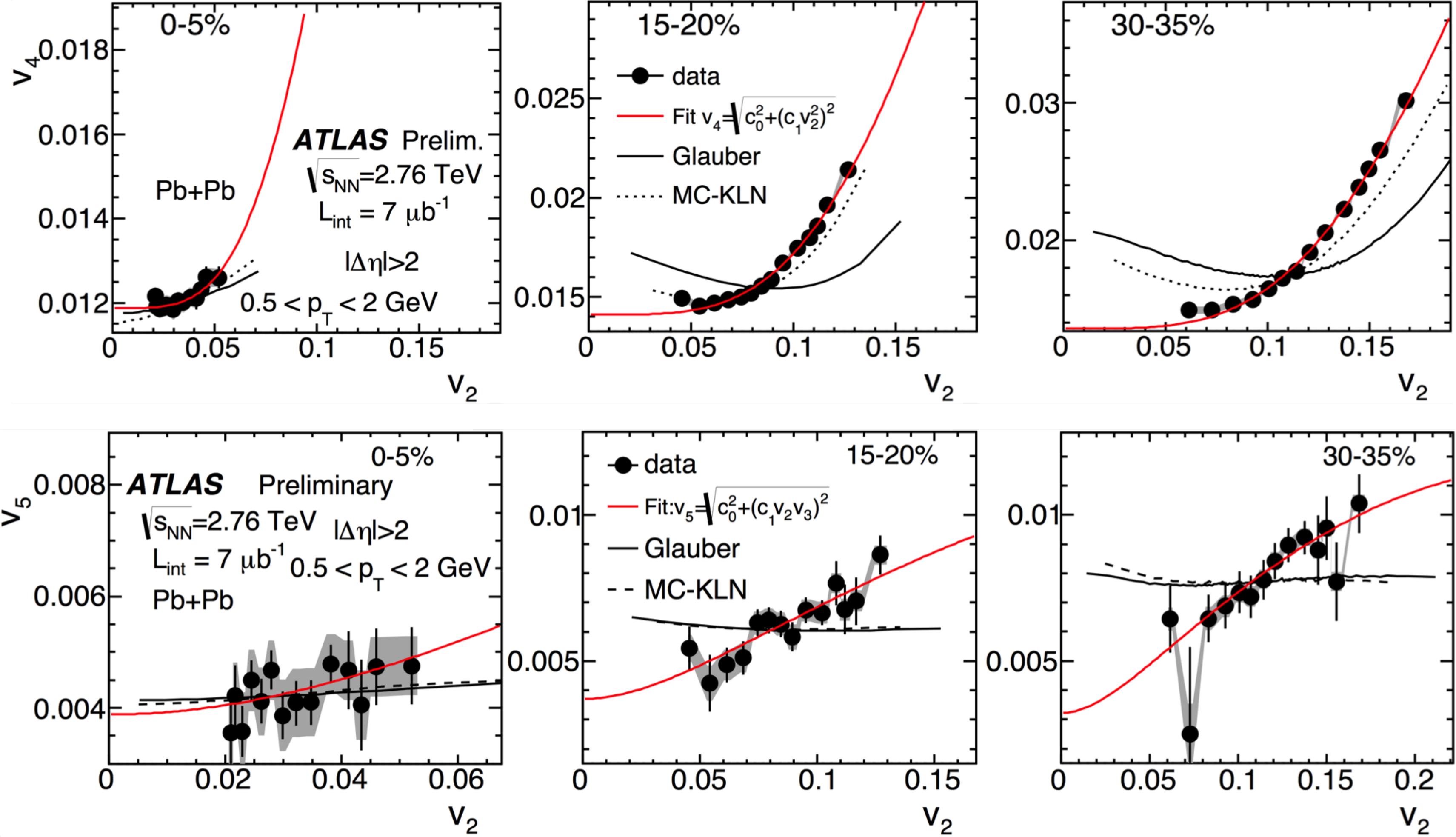}
\caption{\label{fig:res43} The $v_4$--$v_2$ (top row) and $v_5$--$v_2$ (bottom row) correlations measured in $0.5<\pT<2$~GeV in three 5\% centrality intervals. In each panel, the correlation data are fit to functions that include both linear and non-linear contributions. The correlation data are also compared with re-scaled $\varepsilon_n$--$\varepsilon_2$ correlation from the MC Glauber and MC-KLN models in the same centrality interval. Results taken from Ref.~\cite{ATLAS2014-022}.} 
\end{figure}

ATLAS also studied $v_4$--$v_2$ and $v_5$--$v_2$ correlations (see Fig.~\ref{fig:res42}(c) and Fig.~\ref{fig:res43}). The patterns in these correlations are found to be dominated by the interplay between the linear and non-linear collective dynamics in the final state of the Pb+Pb collisions. In fact as shown in Fig.~\ref{fig:res43}, these correlations are well described by two-parameter fits of the following form, motivated by Eq.~\ref{eq:ep3a} for $v_4$ and Eq.~\ref{eq:ep3b} for $v_5$, respectively: 
\begin{eqnarray}
\label{eq:res42}
v_4&=& \sqrt{c_{0}^2 + (c_{1}v_2^2)^{2}}\;,\\
v_5&=& \sqrt{c_{0}^2 + (c_{1}v_2v_3)^{2}}\;,
\end{eqnarray}
These excellent fits suggest that either contributions from higher-order non-linear terms and initial correlation ($\langle\cos4(\Phi_2^*-\Phi_4^*)\rangle$ for $v_4$ and $\langle\cos(2\Phi_2^*+3\Phi_3^*-5\Phi_5^*)\rangle$ or $v_5$) are small, or they are, in effect, included through the non-linear component of the fit.

\begin{figure}
\centering
\includegraphics[width=0.5\linewidth]{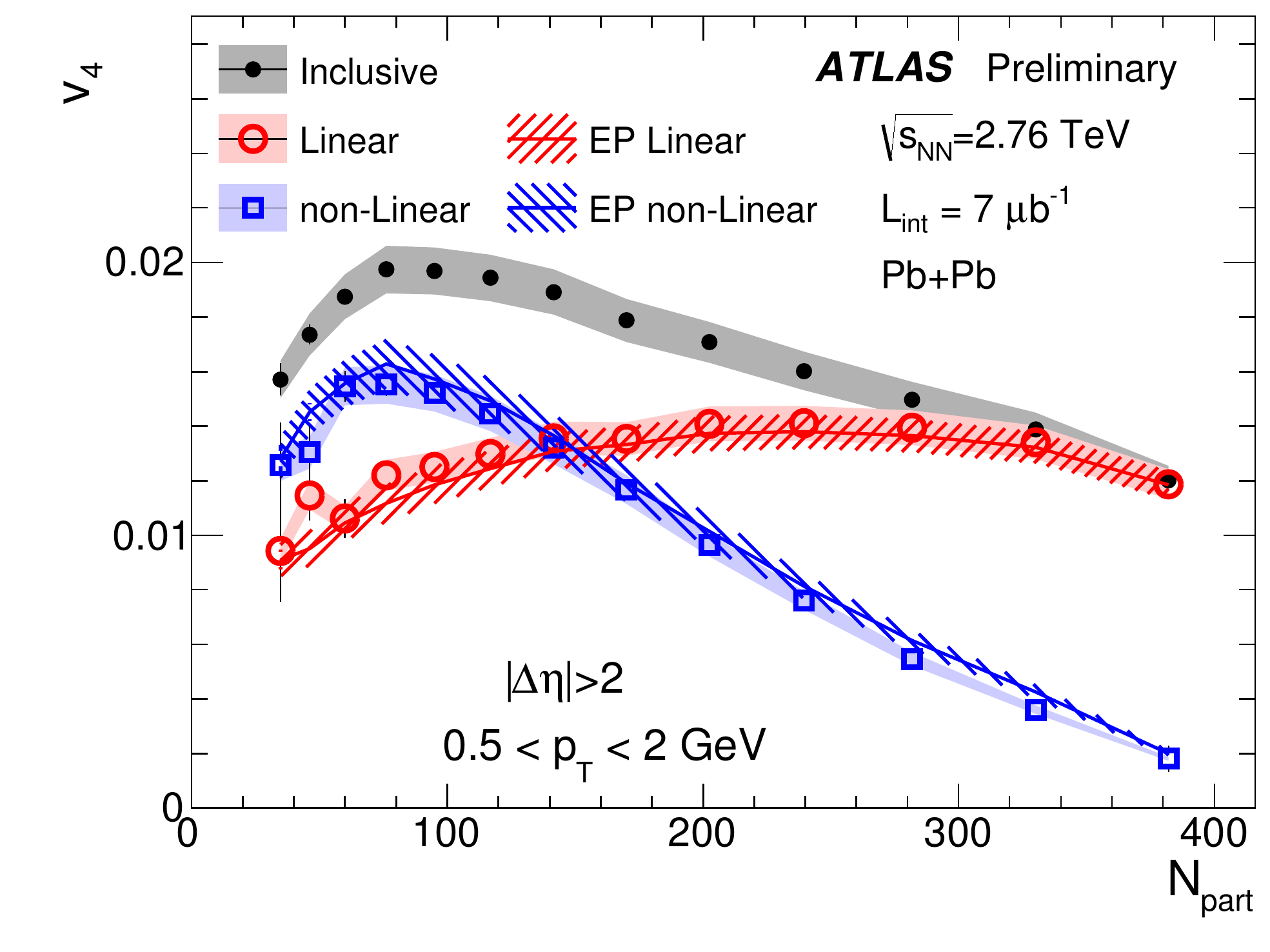}\includegraphics[width=0.5\linewidth]{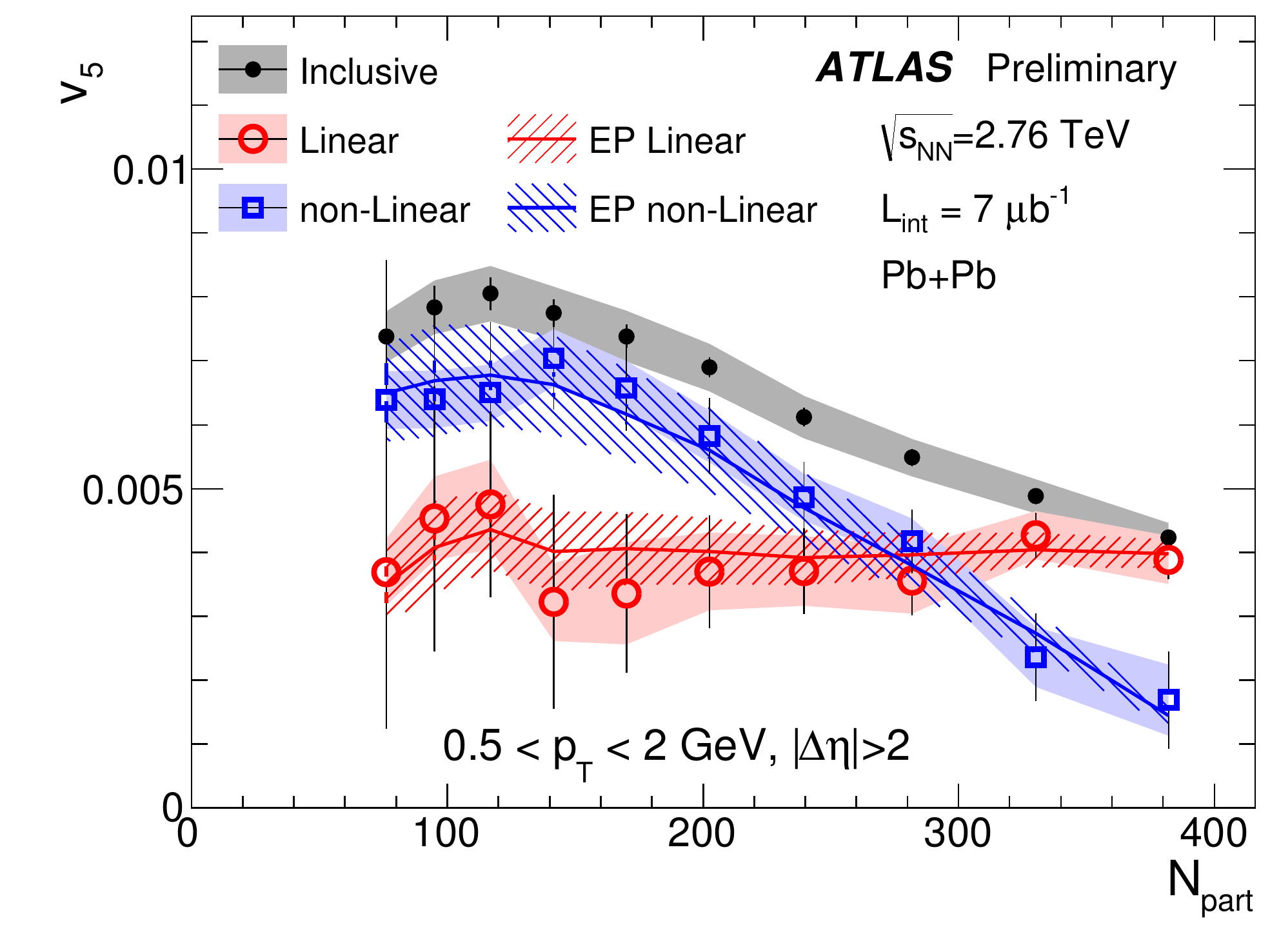}
\caption{\label{fig:res44} The centrality dependence of the $v_4$ (left) and $v_5$ (right) in 0.5--2~GeV and the associated linear and non-linear components extracted from the fit in Fig.~\ref{fig:res43}. They are compared with the linear and non-linear components estimated from the previous published event-plane correlations~\cite{Aad:2014fla}. Results taken from Ref.~\cite{ATLAS2014-022}.}
\end{figure}

The success of the two-component fits naturally allow us to decompose the $v_4$ and $v_5$, centrality by centrality, into linear and non-linear terms as:
\begin{eqnarray}
\label{eq:res43}
v_n^{\rm{L}}=c_0,\;\;\;\; v_n^{\rm{NL}}=\sqrt{v_n^2-c_0^2}
\end{eqnarray}
The results are shown in Fig.~\ref{fig:res44}. The linear term associated with $\varepsilon_n$ depends only weakly on centrality, and dominates the $v_n$ signal in central collisions. The non-linear term increases as the collisions become more peripheral. A similar decomposition can also be obtained directly from the measured event-plane correlations~\cite{Aad:2014fla}:
\begin{eqnarray}
\label{eq:res44}
v_4^{\rm{NL}}=v_4\left\langle\cos4(\Phi_2-\Phi_4)\right\rangle, v_5^{\rm{NL}}=v_5\left\langle\cos(2\Phi_2+3\Phi_3-5\Phi_5)\right\rangle,
\end{eqnarray}
which agree very well with results obtained from direct fits, implying that the correlations between flow magnitudes arise mostly from the correlations between the flow angles.

\subsection{Mixed correlations $p(v_l,\Phi_n,\Phi_m,...)$}
\label{sec:35}
\begin{figure}
\centering
\includegraphics[width=1\linewidth]{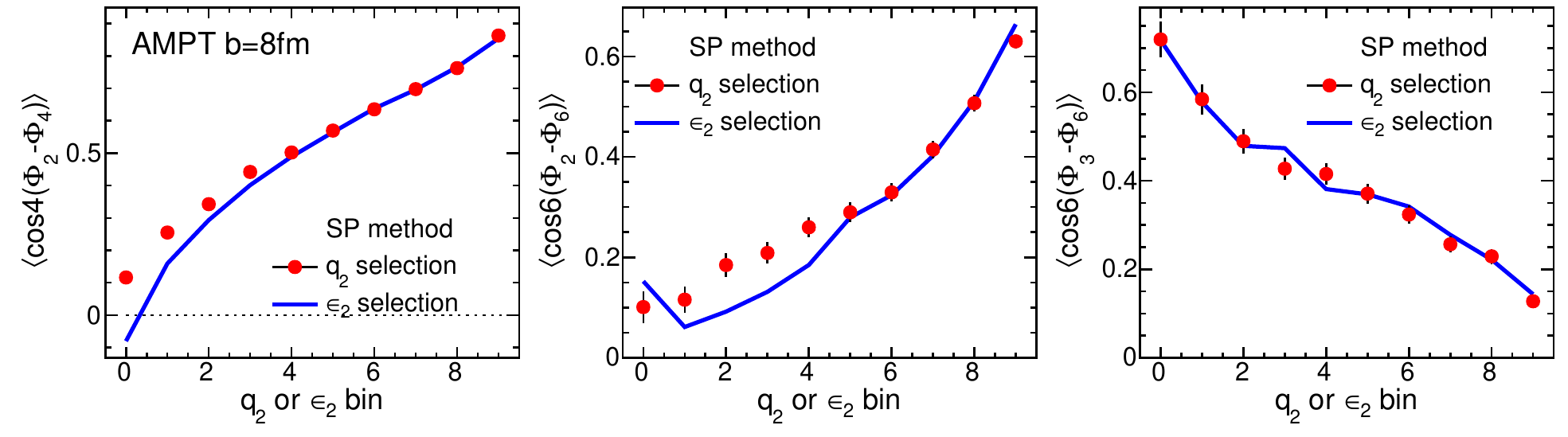}
\caption{\label{fig:res51} Three two-plane correlators in ten bins (each containing 10\% events) of $q_2$ and $\varepsilon_2$ for AMPT Pb+Pb events with $b=8$ fm. Results are calculated using the scalar-production method~\cite{Huo:2013qma}.}
\end{figure}

So far the only study of the correlation between flow magnitudes and event-plane angles is performed in AMPT simulations using the event-shape selection technique~\cite{Huo:2013qma}. In this study, various two- and three-plane correlators are calculated as a function of $v_2$ and $v_3$ for events with the same impact parameter. In many cases, the strength of the event-plane correlation is found to increase with $v_2$ or $v_3$. One example of such study is shown in Fig.~\ref{fig:res51}. The dependence of these correlators on $v_2$ for fixed centrality is qualitatively similar to the overall centrality dependence shown in Fig.~\ref{fig:res31}. That is the correlators $\left\langle\cos 4(\Phi_{2}-\Phi_{4})\right\rangle$ and $\left\langle\cos 6(\Phi_{2}-\Phi_{6})\right\rangle$ increase with $v_2$, while $\left\langle\cos 6(\Phi_{3}-\Phi_{6})\right\rangle$ decreases with $v_2$ (note $v_2$ increases with decreasing $\npart$). These observations imply that most of the mixed-correlations are controlled by the corresponding event-plane correlations, consistent with the results in Sec.~\ref{sec:34}.

\section{Future directions}
\label{sec:4}

\subsection{Precision event-shape selection}
\label{sec:41}

The application of event-shape selection is not only limited to flow observables. It should be straightforward to measure the HBT correlation, nuclear modification $R_{AA}$, dihadron correlation, chiral magnetic effects and more, for events selected to have the same system size but different different ellipticity and triangularity. A recent study from PHENIX collaboration~\cite{esumi} shows that the measured 2$^{\mathrm{nd}}$-order freezeout eccentricity $\varepsilon_2^{\rm f}$ is strongly correlated with event ellipticity. This is qualitatively expected, as events with large $v_2$ should on average have large $\varepsilon_2$, and hence they may have larger $\varepsilon_2^{\rm f}$.

We emphasize that the EbyE fluctuation of initial geometry is very large, and events with the same system size follow very different pathway during the collective expansion. If these events can be classified precisely, one can gain access to huge amount of information on both the initial condition and dynamics in the final state. Studies by the LHC experiments show that the dynamic range for selecting on the average $v_2$ and $\varepsilon_2$ can be as large as a factor of three in mid-central collisions~\cite{ATLAS2014-022,Dobrin:2012zx}. The overlap region of the events with largest $\varepsilon_2$ has a cigar-like shape, with an aspect ratio exceeding a factor of three. Such extremely elongated events are expected to expand voilently along the short-axis direction while exhibiting very little collectivity along the long-axis direction, similar to what is observed of the collective expansion of strongly-interacting cold atoms initially prepared in spatially anisotropic state~\cite{OHara:2002zz}. It would be interesting to perform detailed study of hydrodynamic response for these extreme events. 
 
\subsection{Jet-medium interactions}
\label{sec:42}
One open issue in heavy ion physics is the mechanism of jet-quenching and subsequent dissipation of the lost energy in the medium. The measurement by CMS experiment demonstrates the lost energy of very high $\pT$ jets are transported to very large angle and to low $\pT$ particles~\cite{Chatrchyan:2011sx}. This study is very demanding on the event statistics and the interpretation is complicated by jet selection bias. 

It might be useful to study the fate of the mini-jets that have the energy of a few GeV to few ten's of GeV. These mini-jets are abundantly produced and constitute the bulk of particle spectrum at intermediate $\pT$ where our theoretical understanding is least under control. Traditionally, the medium interaction of mini-jets was accessed using the two-particle azimuthal correlation method, supplemented with the flow background subtraction via ZYAM procedure. This analysis procedure is subject to large systematics due to dominance of collective flow in the correlation structure~\cite{Wang:2013qca}. For example, the away-side double-hump structure in the correlation function after elliptic flow subtraction has been interpreted as the Mach-cone excited by the jets traversing the medium~\cite{Adler:2005ee,Adams:2005ph}, which was latter understood to arise from triangular flow~\cite{Alver:2010gr}.

We may improve the situation by performing the analysis in events with very small $v_n$, taking advantage of the event-shape selection technique. But ultimately, two-particle correlation method may not be the best approach to investigate the jet-medium interactions. Instead, we may need to develop methods that focus directly on the localized $\eta\times\phi$ structures in the EbyE particle multiplicity or $\eT$ distribution.

\begin{figure}
\centering
\includegraphics[width=1\linewidth]{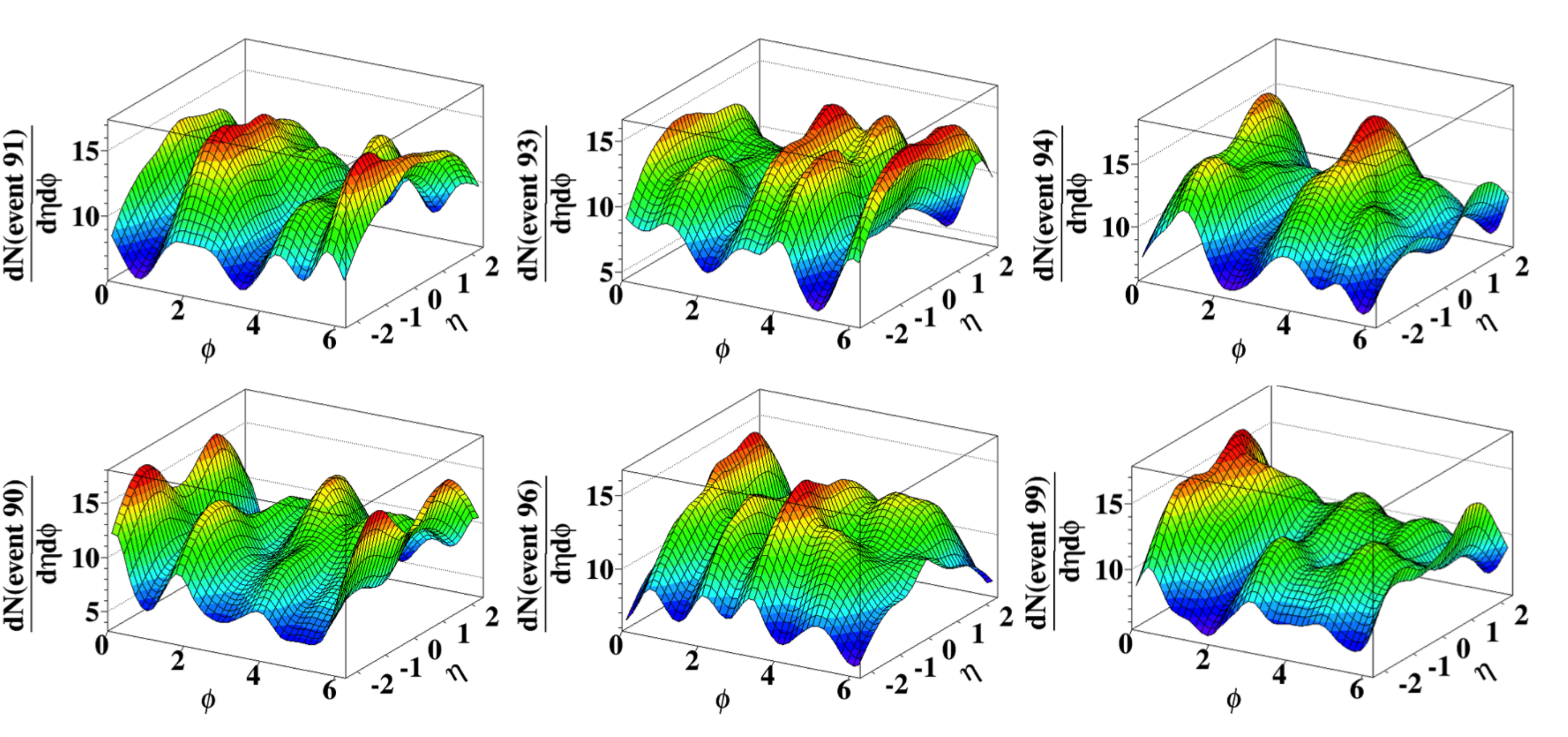}
\caption{\label{fig:pros2} The distributions of the particles in $\eta\times\phi$ space for six typical Au+Au collisions at $\sqrtnn=200$ GeV in 0--10\% centrality interval, obtained from 3+1D calculation based on ideal hydrodynamics~\cite{pang}.}
\end{figure}

To motivate this idea, Figure~\ref{fig:pros2} shows the $\eta\times\phi$ distribution of particle density from a 3+1D EbyE hydrodynamic calculation for several events in 0-10\% based on the AMPT initial condition~\cite{pang}. The localized peaks and valleys in these events could be remnant of the mini-jets in the initial state. These localized structures, or ``hydro-jets'', are much broader than typical high-$\pT$ jets. They can be found, possibly as fake-jets, by running standard jet reconstruction algorithm. Obviously, the jet finding algorithm needs to be modified in order to maximize the finding efficiency and better adapt to the shape of these objects. One can then perform a detailed study of the spectrum and substructure of these hydro-jets.

\subsection{Longitudinal dynamics}
\label{sec:43}
Most studies of the heavy ion collisions only consider fluctuations in the transverse plane, and dynamics in the longitudinal direction are often assumed to be boost invariant. However on an event-by-event basis, there is no apriori reason why this should be the case. Some earlier theoretical works~\cite{Bozek:2010vz,Petersen:2011fp,Xiao:2012uw,Pang:2012uw} show that longitudinal fluctuations may result in non-trivial $\eta$ dependencies of the event-plane angles and two-particle correlation functions. The breaking of boost invariance is especially pronounced for models based on AMPT initial condition, as already shown in Fig.~\ref{fig:pros2}. This section offers some insights on the origin of the longitudinal fluctuations. 

\begin{figure}
\centering
\includegraphics[width=1\linewidth]{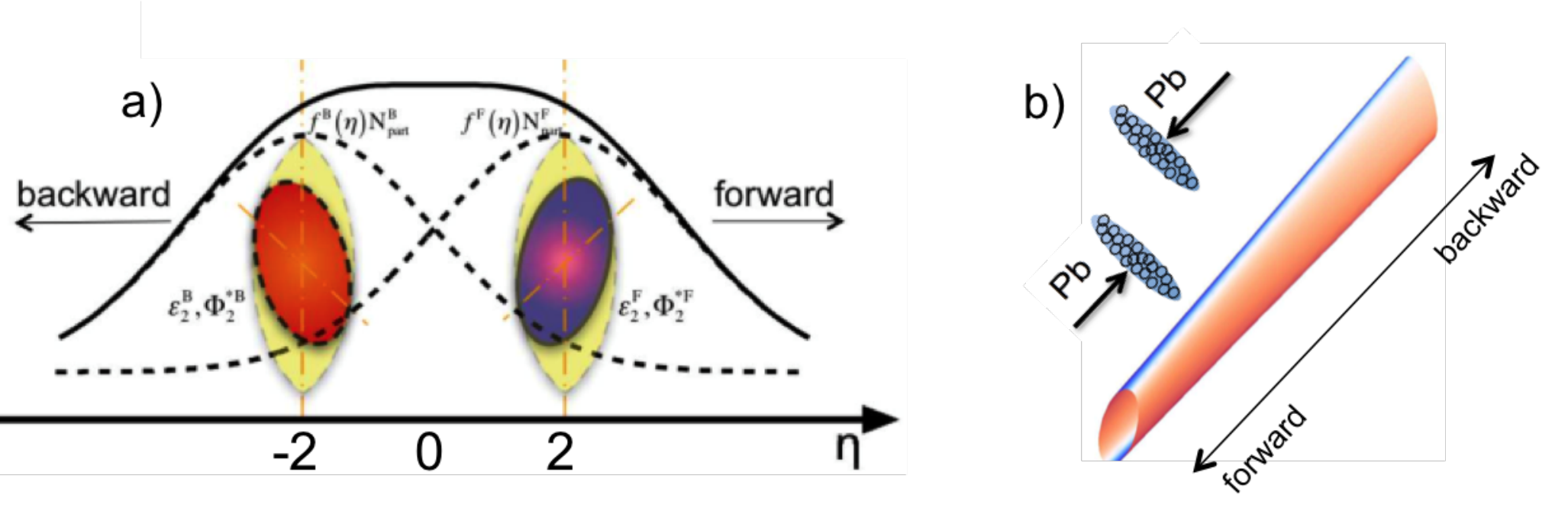}
\caption{\label{fig:pros3}  Schematic illustration of the forward-backward fluctuation of second-order eccentricity and participant plane in A+A collisions. The dashed-lines indicate the particle production profiles for forward-going and backward-going participants, respectively~\cite{Jia:2014vja,Jia:2014ysa,Bozek:2010vz}.}
\end{figure}

We first note that the number of participating nucleons and eccentricity vectors can be separately define for the two colliding nuclei, and in general they can differ strongly on EbyE basis due to fluctuations: $\npartf\neq\npartb$ and $\vec{\varepsilon}_n^{\;\mathrm{F}}\neq \vec{\varepsilon}_n^{\;\mathrm{B}}$. Secondly, the energy deposition for each participating nucleon is not symmetric: Particles in the forward (backward) rapidity are preferably produced by the participants in the forward-going (backward-going) nucleus~\cite{Bialas:1976ed,Bialas:2010zb}. Due to these two effects, the transverse shape of the initially produced fireball at the time of the thermalization but before the onset of the hydrodynamics should be a strong function of $\eta$ (see Fig.~\ref{fig:pros3}). Consequently, the eccentricity vector that drives the evolution of the whole system, $\vec{\varepsilon}_n^{\;\mathrm{tot}}$, is expected to interpolate between $\vec{\varepsilon}_n^{\;\mathrm{F}}$ at forward rapidity and $\vec{\varepsilon}_n^{\;\mathrm{B}}$ at the backward rapidity~\cite{Jia:2014ysa}:
\begin{eqnarray}
\label{eq:pros1}
\vec{\varepsilon}^{\mathrm{\;tot}}_n(\eta)\approx \alpha(\eta)\vec{\varepsilon}_n^{\;\rm F}+(1-\alpha(\eta))\vec{\varepsilon}_n^{\;\rm B}\equiv\varepsilon^{\mathrm{tot}}_n(\eta)e^{in\Phi_n^{*\rm tot}(\eta)}.
\end{eqnarray} 
where $\alpha(\eta)$ is a $\eta$ dependent weighting factor for forward-going participating nucleons. For symmetric collision system and assuming $\npartf=\npartb$, it is approximately $\alpha(\eta) \approx\frac{ f(\eta) }{f(\eta)+f(-\eta)}$, where $f(\eta)$ is the emission profile per-nucleon. Assuming that the harmonic flow at given $\eta$ is driven by the corresponding eccentricity vector at the same $\eta$, which is a reasonable assumption for $n=2$ and 3~\cite{Qiu:2011iv,Gardim:2011xv}, we expect the following relation to hold:
\begin{eqnarray}
\label{eq:pros2}
\vec{v}_n(\eta) &\approx& c_n(\eta)\left[\alpha(\eta)\vec{\varepsilon}_n^{\;\rm F}+(1-\alpha(\eta))\vec{\varepsilon}_n^{\;\rm B}\right]
\end{eqnarray}

\begin{figure}
\centering
\includegraphics[width=1\linewidth]{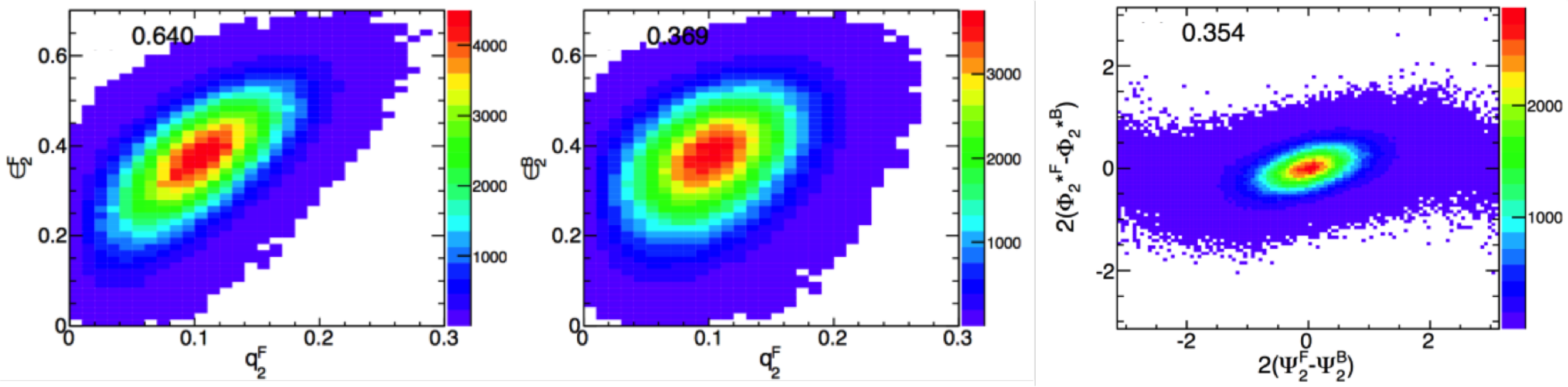}
\caption{\label{fig:pros4} Correlation of $\varepsilon_2^{\mathrm{F}}$ vs. $q_2^{\mathrm{F}}$ (left),  $\varepsilon_2^{\mathrm{B}}$ vs. $q_2^{\mathrm{F}}$ (middle) and initial twist angle vs final state twist angle (right) for AMPT Pb+Pb events with $b=8$~fm. The numbers in first two panels indicate the correlation coefficients~\cite{Jia:2014ysa}.}
\end{figure}

The relation Eq.~\ref{eq:pros2} has been verified using the AMPT model. Figure~\ref{fig:pros4} shows the correlation of the flow vector calculated in a forward $\eta$ range ($4<\eta<6$), $q_2^{\mathrm{F}}$, with the eccentricity for forward-going and backward-going nucleons, $\varepsilon_2^{\mathrm{F}}$ and $\varepsilon_2^{\mathrm{B}}$. The correlation is stronger between $\varepsilon_2^{\mathrm{F}}$ and $q_2^{\mathrm{F}}$ than that between $\varepsilon_2^{\mathrm{B}}$ and $q_2^{\mathrm{F}}$, suggesting that the elliptic flow in the forward-rapidity is driven more by the ellipticity of the forward-going Pb nucleus (and vice versa). Figure~\ref{fig:pros4}(c) shows that the angles between the participant planes are strongly correlated with the angles between the raw event planes, suggesting that the twist in the initial state geometry is converted into twist in the final collective flow between the forward and the backward pseudorapidities. Similar results are also observed for the triangularity and triangular flow.

\begin{figure}
\centering
\includegraphics[width=1\linewidth]{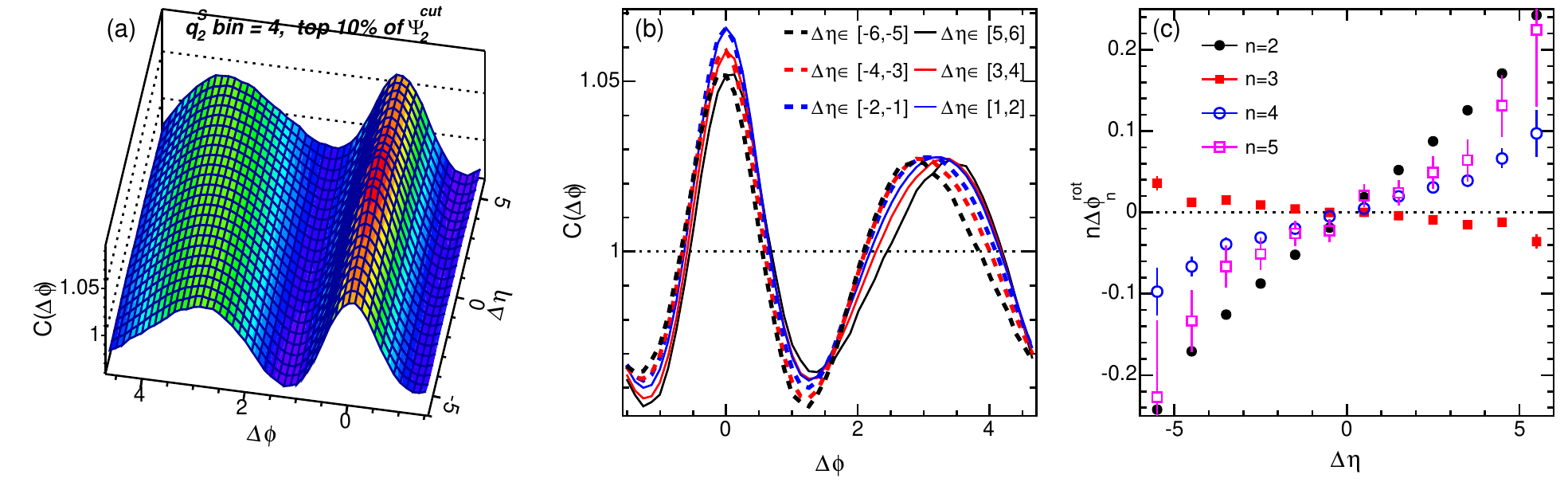}
\caption{\label{fig:pros5} The 2D correlation function (left panel), 1D correlation functions in different $\Deta$ slices (middle panel) and the extracted $n\Phi_n^{\mathrm{rot}}=n(\Phi_n^{\rm F}-\Phi_n^{\rm B})$ for $n=2$--5 (right panel), for events selected in the fourth $q_2^{\rm S}$ bin with largest $\Delta\Psi_2^{\mathrm {cut}}$ values~\cite{Jia:2014vja}.}
\end{figure}

The existence of $\eta$-dependent rotation of the event-plane angle can be unambiguously identified via the event-shape twist method proposed in Ref.~\cite{Jia:2014vja}. In this method, a selection cut is applied on the difference of the event-plane angle between the forward and backward $\eta$ ($4<|\eta|<6$), $\Delta\Psi_2^{\mathrm {cut}}=2(\Psi_2^{\rm F}-\Psi_2^{\rm B})$. A two-particle correlation function is constructed in center-rapidity ($|\eta|<3$) for events selected with large $\Delta\Psi_2^{\mathrm {cut}}$. The effect of twist appears as a twisted-ridge on both the near and away-side as shown in Fig.~\ref{fig:pros5}. In addition, although the selection is enforced for the elliptic flow event plane, a twist is also observed for higher-order harmonics (Fig.~\ref{fig:pros5}(c)). This is characteristic of the effects of non-linear mixing which couples $v_2$ to higher-order $v_n$.  

\section{Summary}

The initial condition of heavy ion collisions is lumpy and fluctuates strongly event to event, each drives an unique collective expansion. Such EbyE fluctuations lead to broad distributions of harmonic flow coefficients $v_n$ and non-trivial correlations between $v_n$ and phases $\Phi_n$ of the harmonic flow. These distributions and correlations lead to a large set of flow observables, which are sensitive to details of the initial collisions geometry and transport properties of medium in the final state. These observables can be measured either directly via a data-driven unfolding method, indirectly by constructing appropriate moments or cumulants using multi-particle correlations, or inferred from event-shape selection techniques. 

Initial measurements of these flow observables have been performed at RHIC and the LHC, and they have been successfully described by sophisticated EbyE hydrodynamic model calculations. The progresses from both fronts help to construct a detailed space-time picture of the heavy ion collision. We now know that the first three flow harmonics are driven mainly by a linear response to the corresponding eccentricity, $v_n\propto\varepsilon_n$ for $n\leq3$. Hence the precision data on $v_2$ and $v_3$, including an apparent anti-correlation between $v_2$ and $v_3$, place a strong constraints on the fluctuations in the initial geometry. For the higher-order flow harmonics $v_4$, $v_5$ and $v_6$, non-linear contributions from lower-order harmonics are very important, especially in mid-central and peripheral collisions. The relative contributions of linear and non-linear effects are sensitive to the expansion dynamics and dictate the experimentally measured correlations between event-plane angles $p(\Phi_n,\Phi_m,...)$ and flow magnitudes $p(v_n,v_m,...)$. In principle, there are enough experimental information to constrain many other important aspects of the heavy ion collisions, such as early thermalization and initial flow, temperature dependence of $\eta/s$, freezeout condition and hadronic transport. However this may be realized only if theoretical models can simultaneously and quantitatively describe all these flow observables.

There are several important open issues that can be addressed in future flow measurements. The event-shape selection technique has been demonstrated to be a very promising tool for elucidating flow response of to the change of initial geometry; More complete picture of system evolution can be obtained by applying this technique to other experimental observables such as HBT, spectra and dihadron correlation measurement. The response or back-reaction of the medium to jets and di-jets is an important component of the flow physics at intermediate and high $\pT$ ($>2$--3 GeV). Measuring such non-equilibrium aspect of the flow physics probably requires going beyond the traditional Fourier-type harmonic analysis and developing new observables that are tuned to the localized structures in $\eta\times\phi$ space of the particle production. Finally, simple arguments based on Glauber model suggest EbyE fluctuations in the longitudinal direction can be as important as the fluctuations in the transverse plane. These initial state longitudinal fluctuations lead to large forward-backward asymmetry in $v_n$ values and twist in the event-plane angle, and are measurable. They promise to open up new avenues for understanding initial state fluctuations, particle production and collective expansion dynamics.

I appreciate fruitful discussions with S.~Mohapatra and comments from J.~Liao, B.~Schenke, R.~Snellings and F.~Wang. This research is supported by NSF under grant number PHY-1305037 and by DOE through BNL under grant number DE-AC02-98CH10886
\label{sec:5}
\section*{References}
\bibliographystyle{atlasBibStyleWoTitle}
\bibliography{flowreview_v4}
\end{document}